\documentclass[
superscriptaddress,
preprint,
 aps,
pra,
]{revtex4-2}
\usepackage[utf8]{inputenc}
\usepackage{amsmath}
\usepackage{amsfonts}
\usepackage{braket}
\usepackage{comment}
\usepackage[normalem]{ulem}
\usepackage{graphicx}
\usepackage{physics}
\usepackage{xcolor}
\usepackage{float}
\usepackage{multirow}
\usepackage{adjustbox}
\usepackage{hyperref}
\usepackage{enumitem,kantlipsum}
\usepackage{upgreek}
\usepackage{soul}

\begin{document}

\preprint{APS/123-QED}
\title{Signum phase mask differential microscopy}
\author{Jeeban Kumar Nayak}
\email{jkn19rs027@iiserkol.ac.in}
\affiliation{Department of Physical Sciences, Indian Institute of Science Education and Research Kolkata, Mohanpur, India, 741246}
\author{Niladri Modak}
\affiliation{Tampere University, Photonics Laboratory, Physics Unit, Tampere, FI-33720, Finland}
\author{Sayan Ghosh}
\affiliation{Department of Physical Sciences, Indian Institute of Science Education and Research Kolkata, Mohanpur, India, 741246}
\author{Nirmalya Ghosh}
\email{nghosh@iiserkol.ac.in}
\affiliation{Department of Physical Sciences, Indian Institute of Science Education and Research Kolkata, Mohanpur, India, 741246}
\begin{abstract}
\noindent
We propose and experimentally demonstrate a differential microscopy method to obtain simultaneous amplitude, phase, and quantitative polarization gradient imaging in a single experimental embodiment. A full-field optical spatial differentiator is achieved in a relatively simple setup by placing a glass cover slip as a Signum phase mask in the Fourier plane of a standard $4-f$ imaging system and accordingly named Signum phase mask differential microscopy. The longstanding requisite of polarized light to obtain the spatial differentiation of the field at the object plane is eliminated in our scheme and, hence, leads to the emergence of quantitative differential polarization contrast imaging by integrating polarization degree of freedom as an additional contrast agent in the framework of differential microscopy. Implementation of the proposed differential imaging scheme in high-resolution microscopy is experimentally demonstrated alongside its functionality for a broad wavelength range. Simultaneous acquisition of differential phase, amplitude, and polarization (anisotropy) gradient imaging in a rather elementary optical setup enables a low-cost multi-functional differential microscopy system that is anticipated to emerge as a revolutionary tool in label-free imaging and optical image processing.
\end{abstract}

\maketitle

Interaction of light with inhomogeneous natural objects provides the intrinsic mechanism of image contrast through the spatial modulation of the amplitude, phase, and polarization of the light field, which is traditionally utilized for optical imaging and microscopy for a wide range of applications in various disciplines \cite{microscopyintroduction,evanko2009milestones,polarisationreview,phaspolarization,strainoptical}. One of the major challenges in the early days of microscopy was to visualize the so-called phase objects (nearly transparent) under a conventional bright-field microscope, as these objects influence the amplitude of the incident light marginally. In such cases, phase and polarization imaging can capture essential morphological details about the internal structure of the object, hence, have paramount importance in label-free microscopy \cite{qpireview,phasecontrastunderstanding,phaseimagingrf1,phaseimagingrf2,phaseimagingrf3}. Many of the natural objects of biological or non-biological origin possess intrinsic polarization anisotropy effects, i.e., birefringence (phase anisotropy) and dichroism (amplitude anisotropy) due to their anisotropic molecular or structural organizations \cite{polarisationreview,chipman2018polarized,goldstein2017polarized,gil2022polarized}. Therefore, quantitative probing of the vectorial transformation of light with desirable spatial resolution provides a wealth of morphological, structural, and functional information on the objects and has been a key tool in biomedical and clinical research \cite{polarisationreview}. On the other hand, in the realm of phase imaging, phase contrast \cite{zernike1934diffraction,zernike1942phase,zernike1942phaseII,burch1942phase,martin1947phase} and differential interference contrast (DIC) \cite{nomarski1954dispositif,franccon1957polarization} are the first-generation techniques, and are still widely adopted to produce contrast images of phase objects such as biological specimen and etc. Owing to their characteristic working principles, some longstanding limitations are associated with these traditional phase and differential microscopy methods, e.g., incapability for quantitative phase imaging, complexity and bulkiness of the setup, polarization dependency, etc. \cite{phasecontrastunderstanding,nomarski1954dispositif}. 
\par
In recent years, there has been a renewed engagement towards the development of next-generation differential microscopy and image processing methods with an aim to achieve versatile and compact optical microscopy systems \cite{kwon2020single,zhou2020flat,kwon2018nonlocal}. 
A wide range of optical systems, starting from reflection in glass surfaces \cite{wang2022photonic,shou2023optical,zhu2019generalized,he2020wavelength,he2020spatial,wang2022brewster} to multi-functional meta-surfaces \cite{kwon2020single,zhou2022fourier,zhou2021two,cordaro2019high,zhu2017plasmonic,zhou2019optical,engay2021polarization,cotrufo2022dispersion}, have been utilized to realize various optical spatial differentiator. On the basis of the corresponding working principles, these recently proposed differential imaging methods can be primarily divided into two types. One, where the spatial differentiation is obtained incorporating the polarization-dependent phase or beam shift and can be considered as a DIC-inspired microscopy technique \cite{kwon2020single,zhou2022fourier,shou2023optical,zhu2019generalized}. In the other approach, the spatial dispersion of the meta-surfaces is appropriately tailored to make them perform as a spatial differentiator \cite{silva2014performing,hwang2016optical,kwon2018nonlocal,cordaro2019high,zhou2020flat,cotrufo2022dispersion}. Although these recently developed spatial differentiaton methods have shown great promises in optical computing and the possible intersection with optical microscopy systems, most of these techniques utilize polarized light to achieve the spatial differentiation\cite{kwon2020single,zhou2022fourier,cotrufo2022dispersion,shou2023optical}. This requisite of differential imaging system makes it impossible to capture simultaneous phase and polarization images of a specimen in a single experimental embodiment. In this context, it is pertinent to note that the traditional polarization microscopic approaches are often severely compromised while probing low level spatial fluctuations of polarization anisotropy effects, which is desirable for most practical applications \cite{polarisationreview}. Thus, encompassing polarization as an contrast mechanism within the framework of conventional differential microscopy (which considers only amplitude and phase) can be a revolutionary tool in the realm of label-free microscopy. 
\par
In this work, we propose and experimentally demonstrate a Signum phase mask differential (SPMD) microscopy scheme, which enables simultaneous amplitude, phase, and quantitative differential polarization imaging/polarization gradient imaging in a simple experimental framework. The proposed microscopy scheme exploits the principle of one-dimensional (1D) Hilbert transformation \cite{davis2000image,davis2002selective}, where a full-field spatial differentiator is obtained by placing a Signum phase mask in the Fourier plane of a conventional $4-f$ imaging setup \cite{goodman1996introduction}. This simple technique enables us to realize a multi-functional differential imaging system built by using just a pair of lenses and a commercially available glass cover slip. Importantly, unlike most of the existing differential microscopy methods \cite{nomarski1954dispositif,zhou2022fourier,kwon2020single}, the proposed spatial differentiator does not require polarized light to obtain the differentiation of the electric field distribution at the object plane. This not only reduces the complexity and bulkiness of the microscopy system but also enables utilizing the polarization of light as an additional intrinsic contrast mechanism in the framework of differential microscopy. As a consequence, two new kinds of imaging systems are emerged in the domain of differential microscopy: polarization-resolved differential imaging and quantitative polarization (anisotropy) gradient imaging. It is worth mentioning that, although the Hilbert transformation has been previously used for image processing, its utilization in the realm of differential microscopy and more specifically its potential for merging the polarization and differential imaging was not touched upon \cite{davis2000image,davis2001fractional}. Usually, the polarization properties of an object are described using the polarization anisotropy parameters (birefringence and dichroism) \cite{polarisationreview,goldstein2017polarized,chipman2018polarized,gil2022polarized}. Being quantitative in nature, our proposed SPMD imaging system can quantify the spatial gradient of these anisotropy parameters throughout a specimen and possess the ability to probe and subsequently produce high contrast images of pure polarization objects even with weak spatial variation of polarization anisotropy effects. The polarization-resolved differential microscopy also holds certain advantages to significantly enhance the image contrast of phase-polarization objects, which will be discussed subsequently. The adaptability of the SPMD imaging scheme in high numerical aperture (NA $\sim0.8$) microscopy is validated alongside its applicability for broad wavelength regions by imaging various biological cells. Demonstration of such a low-cost \textit{multi-functional} setup that can be easily integrated with the conventional commercial bright-field microscope is expected to have a significant impact on the next generation of label-free microscopy and optical image processing techniques.
\par
\begin{figure}[H]
      \centering
      \includegraphics[scale=0.25]{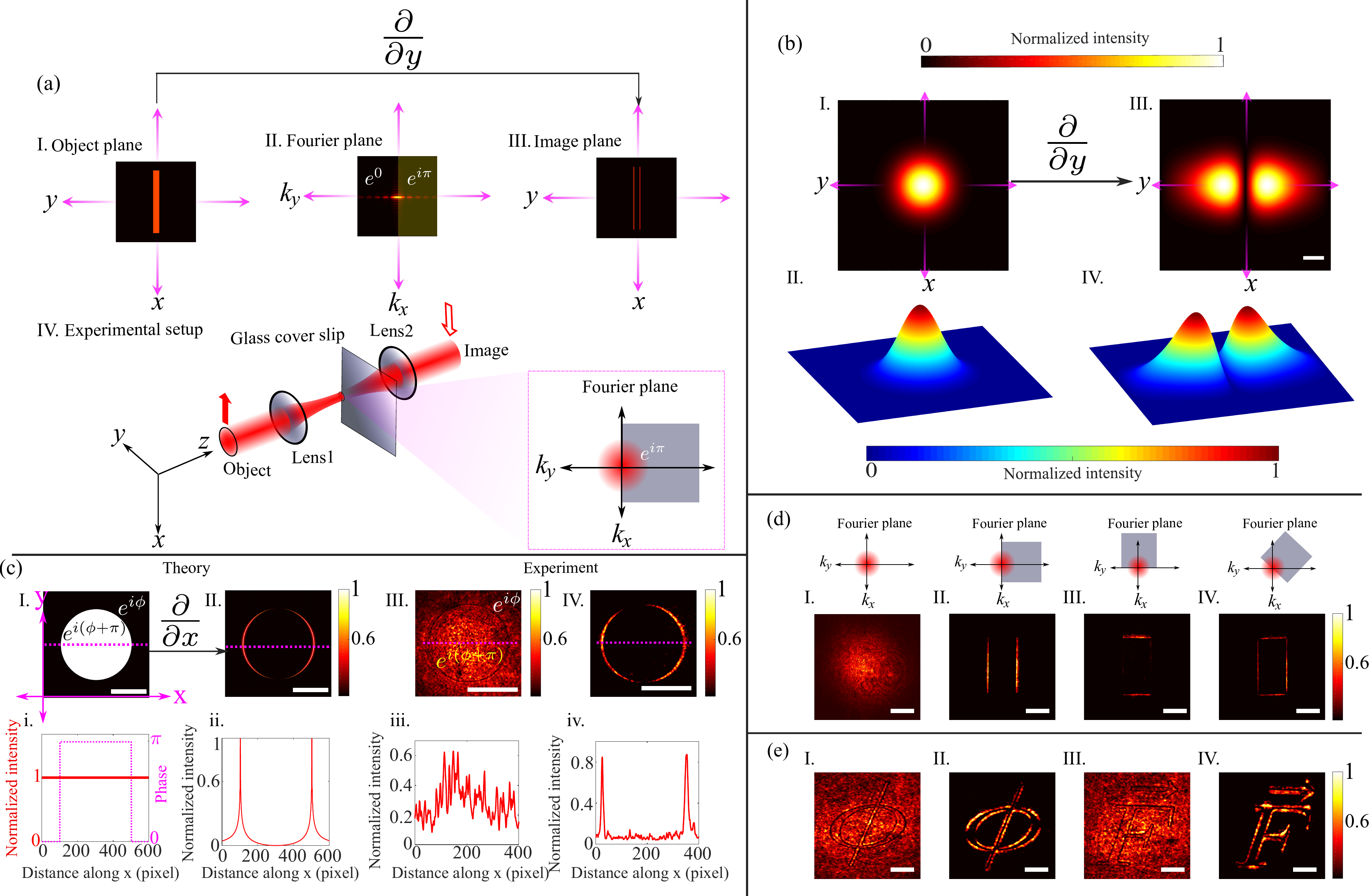}
      \caption{Demonstrating the working principle of the proposed SPMD imaging. (a) $4-f$ experimental arrangement with a $633$ nm laser illumination, where a commercially available glass coverslip (acting as a Signum phase mask) placed in the Fourier plane leads to the formation of the differential image of an input object (I.-IV.). The spatial differentiation scheme is illustrated by considering a rectangular slit as an input object (I.-III.). (b) With an input Gaussian beam (I.), a two-lobed intensity pattern is observed in the output image plane (II.), and the corresponding spatial variation of the intensity profiles (II., IV.) validate formation of differential images with random input objects. (c) Theoretical (I., II., i., ii. ) and experimental (III., IV., iii., iv.) differential images of a circular phase object with a phase discontinuity of $\pi$ at its boundary. The respective intensity profiles (ii., iv.) along the Magenta dashed lines confirm the intensity localizations in the output images at the points of phase discontinuities. (d) Spatial differentiation along a preferred direction is achieved by mechanically rotating the coverslip in the Fourier plane. The differential images of a rectangular phase object (bright field image I.) with spatial differentiation along horizontal $(\hat{x})$ (II.), vertical $(\hat{y})$ (d III.), and $45^{o}$ oriented line (IV.) are presented. (e) Differential images of some additional phase objects prepared by SLM are also shown. All the white scale bars correspond to $1600\mu m$ length ((b)-(e)).}
      \label{fig.1}
\end{figure}

The proposed differential imaging scheme and experimental setup are illustrated in Fig. \ref{fig.1}(a). A commercially available glass cover slip placed in the Fourier plane of a $4-f$ imaging setup imparts different phases $(\phi)$; to the negative $(\phi = \pi: k_y<0)$ and positive $(\phi = 0: k_y>0)$ spatial frequencies on the Fourier plane. Also, due to the high optical scattering at the edge of the cover slip, the transmittance of the zero-order spatial frequencies is reduced to a great extent. In this way, a simple cover slip placed at an appropriate orientation effectively acts as a Signum mask in the Fourier plane \cite{goodman1996introduction}.
The 1D Signum function is chosen for the sake of mathematical simplicity and corresponding experimental feasibility (see Fig. \ref{fig.1}(a)). With a 1D Signum function (in the $y$-direction) modulating the spatial frequencies $(k_x,k_y)$ in the Fourier plane, a full-field spatial differentiator is achieved  where the output intensity distribution at the image plane can be described as (see Supporting information Sec. S.1 for details)
\begin{equation}
\begin{split}
    &\mathbf{E}_{image}(x',y') \propto \int \frac{\partial \mathbf{E}_{obj}(x,y)}{\partial y} \ln |y'-y| dy \\
    &\propto \ln |y'-a_1| + \ln |y'-a_2| + \ln |y'-a_3| +......+ \ln |y'-a_n|
\end{split}
\label{eq1}
\end{equation}
Here $\mathbf{E}_{obj} (x,y)$, describes the spatial variation of the Electric field at the object plane, which corresponds to the complex vectorial transmittance function of the input object, and $y=a_1,a_2,a_3,\ldots,a_n$ are the point of spatial discontinuities/jumps in terms of either the spatial variation of the amplitude, phase, or polarization in the $\mathbf{E}_{obj}$.  As evident from Eq. \eqref{eq1}, the spatial differentiation of the object field distribution $(\mathbf{E}_{obj}(x,y))$ is directly encoded in the captured intensity distribution $(I_{out})$ at the image plane. It is also implied that there will be intensity localization/enhancement at the point of discontinuities in the object field, which leads to the production of high-contrast images and serves as the foundation for the SPMD imaging and microscopy scheme. More importantly, the achieved spatial differentiator considers the vectorial nature of light, which is described by the Jones vector field $[E_1\ E_2]^T$ \cite{polarisationreview,chipman2018polarized}. As a consequence, alongside amplitude and phase objects, differential images of pure polarization objects can be produced in the SPMD imaging scheme by recording the polarization-resolved intensity distribution (Stokes parameters $(S)$ \cite{polarisationreview,gupta2015wave}) at the image plane. The detailed theoretical treatment, particularly, the differential polarization imaging, is provided in the Supporting information Sec. S.1.    
\par
Differential image formation of a rectangular slit utilizing the SPMD imaging scheme is illustrated in Fig. \ref{fig.1}(a). Owing to the amplitude gradient at the boundaries, intensity localization at the edges of the slit is observed at the output image plane. To further validate the proposed spatial differentiation scheme, we numerically simulate the experimental setup and obtain the differential image of an input Gaussian beam (Fig.\ref{fig.1}(b)I., II.), where the spatial intensity distribution at the image plane conforms to the spatial gradient of a Gaussian distribution.  
Next, both numerically and experimentally recorded differential images of a circular phase object are shown Fig. \ref{fig.1}(c). The phase difference between the background and the object leads to an intensity localization at the circumference of the circle (Fig. \ref{fig.1}(c)II., IV.). The extracted intensity profiles in each pixel of the magenta dashed line across both the bright field (Fig. \ref{fig.1}(c)i.,iii.), and differential image (Fig. \ref{fig.1}(c)ii.,iv.) provide compelling proof of the ability of the SPMD imaging scheme to produce high contrast images of pure phase objects. Note that, the spatial variation of the intensity distribution at the image plane (Fig. \ref{fig.1}(c)ii.) emulates the Logarithmic behaviour as obtained in Eq. \eqref{eq1}. Similar differential images of conventional amplitude objects, e.g., slit, and circular aperture, have also been recorded, and corresponding theoretical and experimental results are given in Supporting information Sec. S.2. The direction along which the spatial differentiation is carried out can be conveniently tuned by steering the orientation of the Signum phase mask (glass cover-slip) in the Fourier plane. A rectangular phase object (bright field image: Fig. \ref{fig.1}(d)I.) is utilized to illustrate this, where spatial differentiation along $\hat{y}$ (Fig. \ref{fig.1}(d)II.), $\hat{x}$ (Fig. \ref{fig.1}(d)II.), $\hat{x}+\hat{y}$ (Fig. \ref{fig.1}(d)III.) are performed, and consequent intensity localization along different edges are observed in the corresponding differential images. 
Bright-field and differential images of several phase objects are additionally provided to substantiate the production of high-contrast images of pure phase objects with the SPMD imaging setup (Fig. \ref{fig.1}(e)). A spatial light modulator (SLM) is employed to prepare the phase objects, that are used as imaging targets, and the subsequent differential images are recorded using the SPMD imaging setup (Details regarding optical properties of the used SLM are provided in the method section). 
\par
\begin{figure*}[h!]
      \centering
      \includegraphics[scale=0.3]{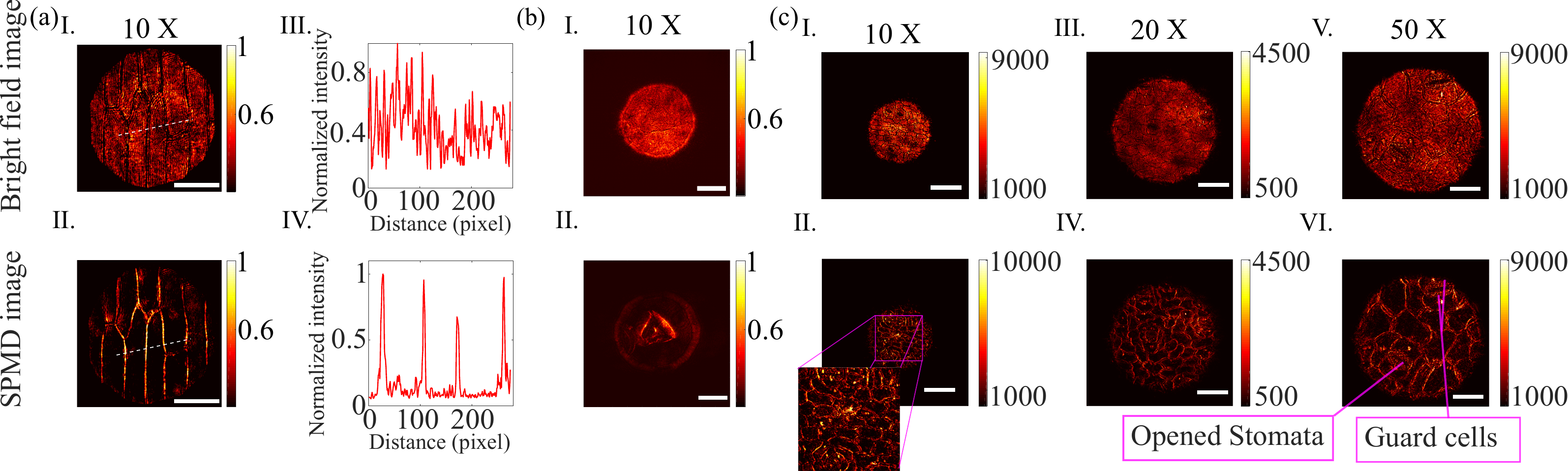}
      \caption{Imaging unlabeled biological cells using the SPMD microscopy. Bright field (top row), and the corresponding differential images (bottom row) are recorded using a $633$ nm laser illumination source. (a) Onion cell (bright field image I.) imaged by the SPMD microscope (II.) with 10X objective shows contrast improvement that is apparent also from the obtained intensity profiles (III.,IV.) white dotted lines of I.,II. (b) Similarly, the invisible cell structures, i.e., cell boundary and nucleus, of a human cheek cell (bright field image I.) are revealed in the obtained differential images (II.) (c) Validating the performance of high numerical aperture (NA) imaging using the SPMD microscope. The cross-section of a plant (\textit{Aglaonema commutatum}) leaf is imaged in the SPMD microscope using the objective lenses of different magnifications/NA (II. 10X $\approx$ 0.25, IV. 20X $\approx$ 0.4, VI. 50X $\approx$ 0.8). Corresponding bright field images are presented in I., III., and V. The clearly visible stomata and guard cells (spatial dimension $\sim 50\mu m$) in the captured differential images demonstrate the formation of high-contrast images with high spatial resolution in the proposed SPMD microscopy. White scale bars correspond to $1600\mu m$ length}
      \label{fig.2}
\end{figure*}

Next, we demonstrate the adaptability of the SPMD scheme in the domain of microscopy. To do so, commercial microscope objectives are integrated with the existing spatial differentiator setup and various biological cells are used as imaging specimens (Fig. \ref{fig.2}). While the bright-field images of the unstained onion and human cheek cells show low contrast due to their pure phase nature (Fig. \ref{fig.2}(a)I.,(b)I.), corresponding differential images exhibit high contrast with clearly visible cell structures (Fig. \ref{fig.2}(a)II.,(b)II)(intensity profile corresponding to the white dashed line is given in (Fig. \ref{fig.2}(a)III.,IV). Moreover, we have recorded the differential images of an \textit{Aglaonema commutatum} leaf cell using three different microscope objectives with 10X (numerical aperture, NA $\approx 0.3$), 20X (NA $\approx 0.4$), and 50X (NA $\approx 0.8$) magnification (Fig.\ref{fig.2}c). The observed differential images authenticate the ability of the SPMD microscope to produce high-contrast images even with high spatial resolution. Thus, it is evident that the proposed SPMD microscopy scheme holds great potential for real-time label-free imaging of transparent biological objects. In addition to this, we also note that the spatial differentiation process in the SPMD scheme is intrinsically independent of the wavelength of the incident light, and thus, the same experimental arrangement can be utilized for a broad range of wavelengths. Indeed, high-contrast differential images of pure phase objects are obtained with laser illumination at different wavelengths,  keeping the optical spatial differentiator unchanged (see Sec S.5 of SI).
\par
\begin{figure}[h!]
      \centering
      \includegraphics[scale=0.3]{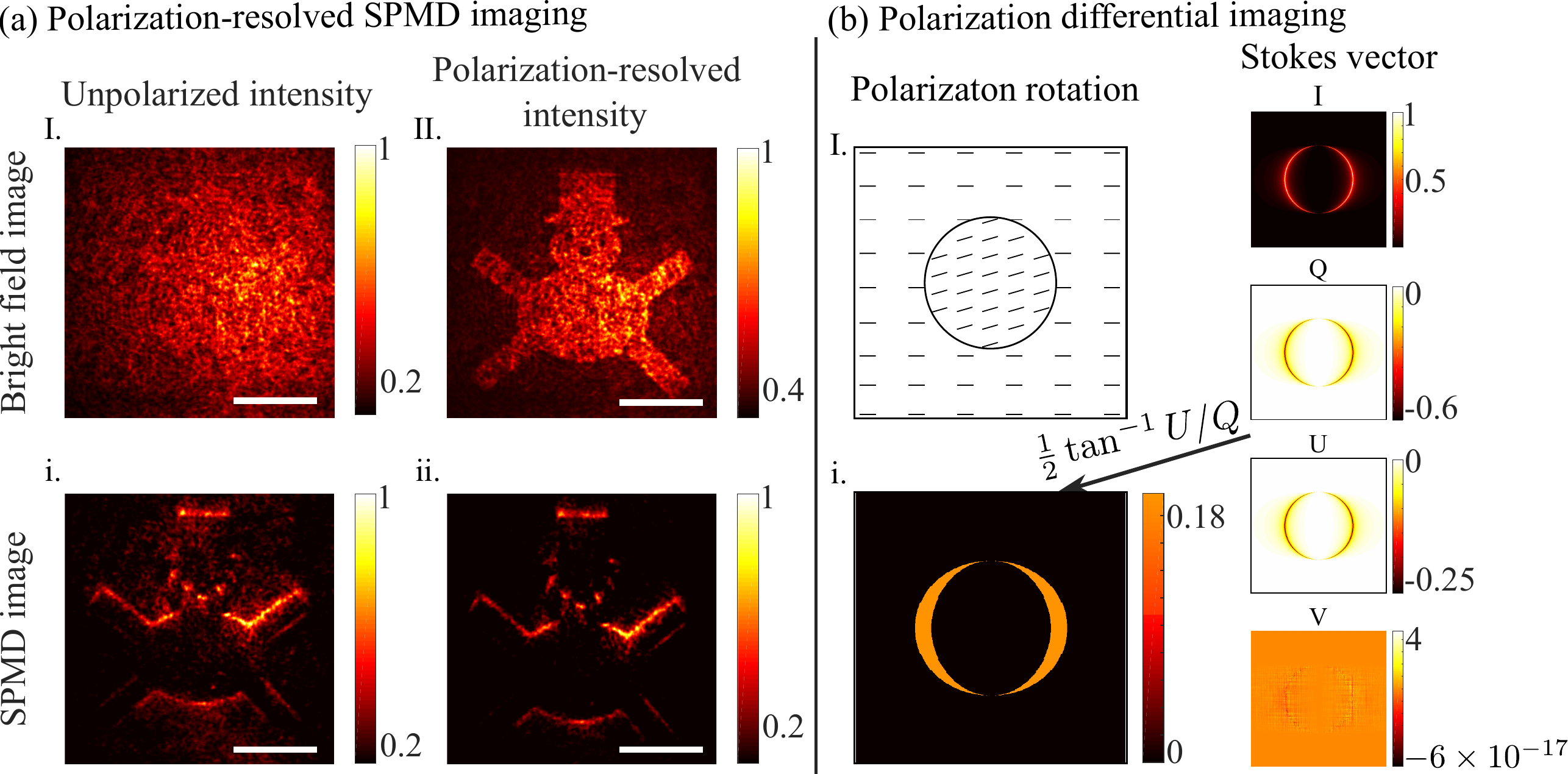}
      \caption{Demonstration of the (a) polarization-resolved differential imaging, and (b) polarization gradient imaging in the SPMD imaging set up. (a) A phase-polarization object, shaped as a snowman, prepared using the SLM, is used as an imaging target. Both the bright field (I.,II.), and differential (i.,ii) images of the input object are recorded with unpolarized (I.,i.) light, and also with the polarization projective measurements (II.,ii.). The polarization-resolved bright field (II.) and differential (ii.) images exhibit significantly greater contrast with respect to their unpolarized counterparts. (b) Differential image of a circle-shaped pure polarization object is presented, where the polarization inside a circle is slightly rotated (due to the circular retardance effect) with respect to the background's polarization. The discontinuity in the polarization anisotropy parameter leads to an intensity localization around the circumference of the circle (Stokes parameter I). The corresponding Stokes parameters and the calculated polarization rotation (i.) demonstrate the quantitative polarization gradient imaging. The white scale bars correspond to $800\mu m$ spatial distance.}
      \label{fig.3}
\end{figure}

As discussed earlier, one of the biggest advantages of the SPMD imaging scheme is that it can utilize the polarization degree of freedom as an additional contrast mechanism in the domain of differential microscopy. We first demonstrate, how the integration of polarization-resolved measurements with differential imaging can significantly enhance the image contrast. For this purpose, a phase-polarization object is prepared using the SLM, and the polarization-resolved intensity patterns at the output image plane are recorded by inserting conventional polarizing optical elements, such as linear polarizers and quarter wave plates (Fig. \ref{fig.3}(a)). Even in the case of bright-field imaging, the polarization-resolved measurement (Fig. \ref{fig.3}(a)II.) reveals the structure of the object (by reducing the overwhelmed background signal), which is not visible in the unpolarized intensity (Fig. \ref{fig.3}(a)I.). Consequently, the differential image corresponding to the polarization-resolved intensity pattern (Fig. \ref{fig.3}(a)ii.) exhibits much higher contrast as compared to that with the unpolarized light (Fig. \ref{fig.3}(a)i.). This observation highlights the utilization of polarization as an additional handle to improve the image contrast in SPMD microscopy.
\par 
Next, we describe the quantitative polarization gradient imaging capability of the proposed SPMD scheme. As Eq. \eqref{eq1} suggests, the proposed spatial differentiator can transfer the information of polarization anisotropy \cite{modak2021generalized} gradient at the object plane to the polarization-resolved intensity distribution at the image plane. This is illustrated considering a pure polarization object, where although there is no change in the absolute amplitude or phase of the electric field transmitted through the object $\mathbf{E_{obj}}$, the polarization of the transmitted light gets modified due to the anisotropic nature of the object, and differ with respect to the background signal. The effect of such discontinuity in the polarization anisotropy parameter, e.g., circular retardance (optical rotation) around the circumference of a circle-shaped object, is studied (Fig. \ref{fig.3}(b)I.). As a consequence of the polarization discontinuity, differential polarization-contrast images are obtained by recording the Stokes vector elements ($[I\ Q\ U\ V]^T$) at the image plane (Fig. \ref{fig.3}(b)I.). The spatial gradient of the polarization anisotropy effect (rotation) is also quantified using standard polarization algebra $(\Psi=atan(U/Q))$ and presented in Fig. \ref{fig.3}(b)i. (see  Supporting information Sec. S.1 for details). In the case of natural polarization objects, that exhibit multiple anisotropy effects simultaneously, such as biological cells and tissues \cite{polarisationreview}, recording the $4\times4$ Mueller matrix (in the same experimental set up- see Supporting information Sec. S.4) and subsequent Mueller matrix decomposition analysis can provide a complete quantitative polarimetric information throughout the specimen. More importantly, this kind of polarization differential microscopy can be advantageous while investigating the morphology of an object with weak spatial variation of polarization anisotropy effects, thus showing promise in biomedical and clinical research.
\par
\begin{figure}[h!]
      \centering
      \includegraphics[scale=0.35]{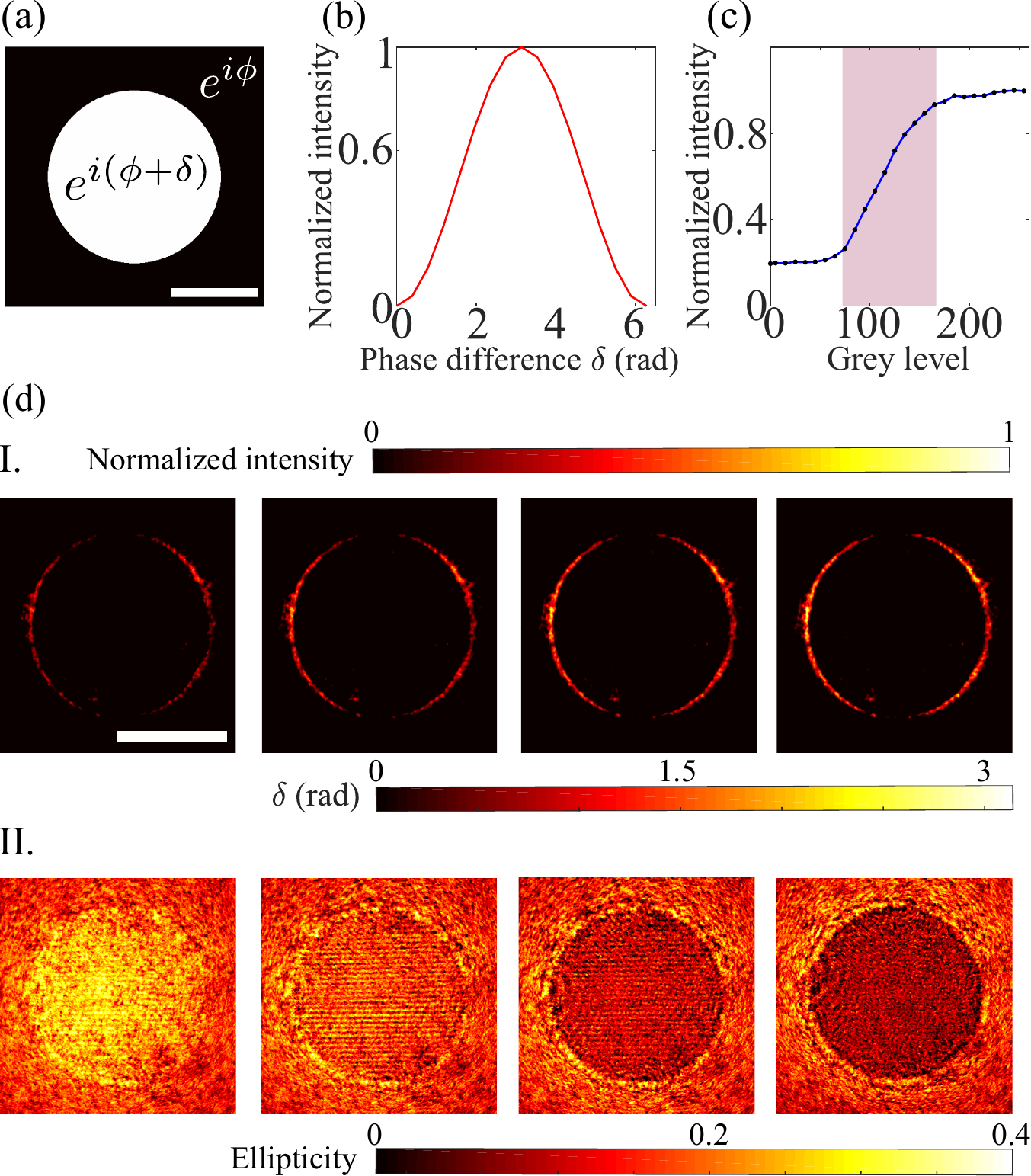}
      \caption{Demonstration of simultaneous quantitative phase and polarization imaging using the SPMD imaging setup. (a) A circular phase object with a phase change of $\delta$ with respect to the background is considered. (b) The theoretical results describing the variation of localized intensity with $\delta$ from $0$ to $\pi$ is plotted (corresponding differential images are presented in Sec. S.3 of SI) (c) The corresponding experimental results are obtained using the SLM, where the full dynamic range of the SLM is used. The reported variation of phase with changing grey levels\cite{pal2016tunable} is exploited to establish a direct relation between the phase difference and intensity localization at the circumference. (d) The corresponding quantified phase images of the circle are presented. (e) The quantitative polarization-resolved measurement is carried out to find the linear retardance parameter $(\alpha)$ corresponding to different grey values. The white scale bar corresponds to $1600 \mu m$ spatial distance.}
      \label{fig.4}
\end{figure}

We now shift our discussion to the possible quantitative phase imaging using the SPMD scheme. Similar to the conventional differential imaging methods, in the SPMD imaging scheme, the spatial field gradient at the object plane dictates the intensity distribution at the image plane. Although the theoretical expression obtained for the output intensity pattern $(I_{out})$ implies a blow-up of intensity values at the point of field discontinuity (Eq. \ref{eq1}), it is never the case for practical situations due to the non-ideal behaviour of real step functions, and its corresponding derivatives. Thus, the magnitude of the intensity distribution at the image plane can be incorporated to quantitatively extract the spatial phase gradient distribution, which is illustrated next.
The behaviour of the localized intensity values with different phase gradients is both numerically (Fig. \ref{fig.4}(b)) calculated and experimentally (Fig. \ref{fig.4}(c)) studied in a circular phase object, where the phase difference $\delta$ between the object and the background is increased gradually. In both cases, the maximum magnitude of the localized intensity at the boundary (edge of the circle) exhibits a monotonically increasing behavior for $\delta = 0$ to $\pi$ (Fig. \ref{fig.4}(b), (c)). This indicates the possibility of establishing a direct relation between the phase gradient and the magnitude of the localized intensity, which can be utilized to quantify the local spatial phase gradient itself (see  Supporting information Sec. S.3). To experimentally demonstrate the quantitative phase imaging capability of the SPMD scheme, circular phase objects are prepared using the SLM with varying phases (achieved by varying the grey levels). The phase step is gradually increased, which is manifested in the maxima of the localized intensities in the corresponding differential images (Fig. \ref{fig.4}(d)I.). Now, the priorly obtained relation (calibration) between the phase gradient and the magnitude of the localized intensity (Fig. \ref{fig.4}(c) is exploited to quantify the phase gradient at the boundaries of the objects, and consequently, the phase distributions of the circular objects are derived (see  Supporting information Sec. S.3). Also, the utilized SLM in the experiment modulates the polarization properties (linear retardance) of light (hence the projected object), and quantitative polarimetric imaging of the object is also obtained by recording the polarization-resolved Stokes vectors (Fig. \ref{fig.4}(d)II.). Through this illustration, an imaging system with the ability of simultaneous quantitative phase and polarization imaging is realized in a single experimental embodiment.
\par
In summary, we have demonstrated a Signum phase mask differential (SPMD) imaging and microscopy protocol with the capability of simultaneous differential phase, differential amplitude, and quantitative polarization gradient imaging in a simple experimental embodiment. The longstanding requisite of using polarization degree of freedom of light to obtain a spatial differentiation of image is eliminated in our scheme, enabling us to integrate polarization as a contrast mechanism in the framework of differential imaging. Importantly, the SPMD imaging scheme leads to a novel quantitative polarization gradient imaging system with the potential to probe low levels and small spatial variations of polarization anisotropy in a sample. The utilization of the proposed SPMD scheme in the domain of microscopy is validated by producing high contrast images of biological samples with high spatial resolution by integrating high numerical aperture objectives. Simultaneous demonstration of amplitude, phase, and polarization contrast differential microscopy in a simple experimental embodiment, imaging capability with high spatial resolution, functionality over a broad wavelength region, quantitative and semi-quantitative extraction of polarization and phase gradient information, all together pave the way for new generation cost-effective multi-functional microscopy in label-free imaging and image processing.

\section*{Methods: Preparation of phase and polarization objects using a spatial light modulator.}
The spatial light modulator (SLM) is generally used to modulate the amplitude, phase, and polarization of light in space and time \cite{efron1994spatial}. Our work uses a twisted nematic liquid crystal-based SLM \cite{pal2016tunable} to generate several phase and polarization objects that are used as imaging targets in the SPMD imaging setup. A specific grey level value in the SLM represents a defined average voltage across the liquid crystal cell. This voltage leads to a variable tilt of the anisotropy liquid crystal molecules. Due to the inherent anisotropy properties of the liquid crystal molecules, the SLM induces a particular phase according to the tilt angle of the liquid crystal layer. In the case of our twisted SLM, the net anisotropy effects are manifested as linear/circular retardance and the accumulation of dynamic phases. The complete polarization and phase modulation properties of the utilized SLM were obtained through rigorous calibration \cite{pal2016tunable}. Consequently, the accumulated phase and polarization effects as the function of grey levels are obtained, which are used to prepare regulated phase and polarization objects.
\bibliography{apssamp}

\begin{thebibliography}{46}%
\makeatletter
\providecommand \@ifxundefined [1]{%
 \@ifx{#1\undefined}
}%
\providecommand \@ifnum [1]{%
 \ifnum #1\expandafter \@firstoftwo
 \else \expandafter \@secondoftwo
 \fi
}%
\providecommand \@ifx [1]{%
 \ifx #1\expandafter \@firstoftwo
 \else \expandafter \@secondoftwo
 \fi
}%
\providecommand \natexlab [1]{#1}%
\providecommand \enquote  [1]{``#1''}%
\providecommand \bibnamefont  [1]{#1}%
\providecommand \bibfnamefont [1]{#1}%
\providecommand \citenamefont [1]{#1}%
\providecommand \href@noop [0]{\@secondoftwo}%
\providecommand \href [0]{\begingroup \@sanitize@url \@href}%
\providecommand \@href[1]{\@@startlink{#1}\@@href}%
\providecommand \@@href[1]{\endgroup#1\@@endlink}%
\providecommand \@sanitize@url [0]{\catcode `\\12\catcode `\$12\catcode `\&12\catcode `\#12\catcode `\^12\catcode `\_12\catcode `\%12\relax}%
\providecommand \@@startlink[1]{}%
\providecommand \@@endlink[0]{}%
\providecommand \url  [0]{\begingroup\@sanitize@url \@url }%
\providecommand \@url [1]{\endgroup\@href {#1}{\urlprefix }}%
\providecommand \urlprefix  [0]{URL }%
\providecommand \Eprint [0]{\href }%
\providecommand \doibase [0]{https://doi.org/}%
\providecommand \selectlanguage [0]{\@gobble}%
\providecommand \bibinfo  [0]{\@secondoftwo}%
\providecommand \bibfield  [0]{\@secondoftwo}%
\providecommand \translation [1]{[#1]}%
\providecommand \BibitemOpen [0]{}%
\providecommand \bibitemStop [0]{}%
\providecommand \bibitemNoStop [0]{.\EOS\space}%
\providecommand \EOS [0]{\spacefactor3000\relax}%
\providecommand \BibitemShut  [1]{\csname bibitem#1\endcsname}%
\let\auto@bib@innerbib\@empty
\bibitem [{\citenamefont {Mertz}(2019)}]{microscopyintroduction}%
  \BibitemOpen
  \bibfield  {author} {\bibinfo {author} {\bibfnamefont {J.}~\bibnamefont {Mertz}},\ }\href@noop {} {\emph {\bibinfo {title} {Introduction to optical microscopy}}}\ (\bibinfo  {publisher} {Cambridge University Press},\ \bibinfo {year} {2019})\BibitemShut {NoStop}%
\bibitem [{\citenamefont {Evanko}\ \emph {et~al.}(2009)\citenamefont {Evanko}, \citenamefont {Heinrichs},\ and\ \citenamefont {Rosenthal}}]{evanko2009milestones}%
  \BibitemOpen
  \bibfield  {author} {\bibinfo {author} {\bibfnamefont {D.}~\bibnamefont {Evanko}}, \bibinfo {author} {\bibfnamefont {A.}~\bibnamefont {Heinrichs}},\ and\ \bibinfo {author} {\bibfnamefont {C.}~\bibnamefont {Rosenthal}},\ }\bibfield  {title} {\bibinfo {title} {Milestones in light microscopy},\ }\href@noop {} {\bibfield  {journal} {\bibinfo  {journal} {Nature Cell Biol}\ }\textbf {\bibinfo {volume} {11}},\ \bibinfo {pages} {S5} (\bibinfo {year} {2009})}\BibitemShut {NoStop}%
\bibitem [{\citenamefont {He}\ \emph {et~al.}(2021)\citenamefont {He}, \citenamefont {He}, \citenamefont {Chang}, \citenamefont {Chen}, \citenamefont {Ma},\ and\ \citenamefont {Booth}}]{polarisationreview}%
  \BibitemOpen
  \bibfield  {author} {\bibinfo {author} {\bibfnamefont {C.}~\bibnamefont {He}}, \bibinfo {author} {\bibfnamefont {H.}~\bibnamefont {He}}, \bibinfo {author} {\bibfnamefont {J.}~\bibnamefont {Chang}}, \bibinfo {author} {\bibfnamefont {B.}~\bibnamefont {Chen}}, \bibinfo {author} {\bibfnamefont {H.}~\bibnamefont {Ma}},\ and\ \bibinfo {author} {\bibfnamefont {M.~J.}\ \bibnamefont {Booth}},\ }\bibfield  {title} {\bibinfo {title} {Polarisation optics for biomedical and clinical applications: a review},\ }\href@noop {} {\bibfield  {journal} {\bibinfo  {journal} {Light: Science \& Applications}\ }\textbf {\bibinfo {volume} {10}},\ \bibinfo {pages} {194} (\bibinfo {year} {2021})}\BibitemShut {NoStop}%
\bibitem [{\citenamefont {Guo}\ \emph {et~al.}(2020)\citenamefont {Guo}, \citenamefont {Yeh}, \citenamefont {Folkesson}, \citenamefont {Ivanov}, \citenamefont {Krishnan}, \citenamefont {Keefe}, \citenamefont {Hashemi}, \citenamefont {Shin}, \citenamefont {Chhun}, \citenamefont {Cho} \emph {et~al.}}]{phaspolarization}%
  \BibitemOpen
  \bibfield  {author} {\bibinfo {author} {\bibfnamefont {S.-M.}\ \bibnamefont {Guo}}, \bibinfo {author} {\bibfnamefont {L.-H.}\ \bibnamefont {Yeh}}, \bibinfo {author} {\bibfnamefont {J.}~\bibnamefont {Folkesson}}, \bibinfo {author} {\bibfnamefont {I.~E.}\ \bibnamefont {Ivanov}}, \bibinfo {author} {\bibfnamefont {A.~P.}\ \bibnamefont {Krishnan}}, \bibinfo {author} {\bibfnamefont {M.~G.}\ \bibnamefont {Keefe}}, \bibinfo {author} {\bibfnamefont {E.}~\bibnamefont {Hashemi}}, \bibinfo {author} {\bibfnamefont {D.}~\bibnamefont {Shin}}, \bibinfo {author} {\bibfnamefont {B.~B.}\ \bibnamefont {Chhun}}, \bibinfo {author} {\bibfnamefont {N.~H.}\ \bibnamefont {Cho}}, \emph {et~al.},\ }\bibfield  {title} {\bibinfo {title} {Revealing architectural order with quantitative label-free imaging and deep learning},\ }\href@noop {} {\bibfield  {journal} {\bibinfo  {journal} {elife}\ }\textbf {\bibinfo {volume} {9}},\ \bibinfo {pages} {e55502} (\bibinfo {year} {2020})}\BibitemShut {NoStop}%
\bibitem [{\citenamefont {Mennel}\ \emph {et~al.}(2018)\citenamefont {Mennel}, \citenamefont {Furchi}, \citenamefont {Wachter}, \citenamefont {Paur}, \citenamefont {Polyushkin},\ and\ \citenamefont {Mueller}}]{strainoptical}%
  \BibitemOpen
  \bibfield  {author} {\bibinfo {author} {\bibfnamefont {L.}~\bibnamefont {Mennel}}, \bibinfo {author} {\bibfnamefont {M.~M.}\ \bibnamefont {Furchi}}, \bibinfo {author} {\bibfnamefont {S.}~\bibnamefont {Wachter}}, \bibinfo {author} {\bibfnamefont {M.}~\bibnamefont {Paur}}, \bibinfo {author} {\bibfnamefont {D.~K.}\ \bibnamefont {Polyushkin}},\ and\ \bibinfo {author} {\bibfnamefont {T.}~\bibnamefont {Mueller}},\ }\bibfield  {title} {\bibinfo {title} {Optical imaging of strain in two-dimensional crystals},\ }\href@noop {} {\bibfield  {journal} {\bibinfo  {journal} {Nature communications}\ }\textbf {\bibinfo {volume} {9}},\ \bibinfo {pages} {516} (\bibinfo {year} {2018})}\BibitemShut {NoStop}%
\bibitem [{\citenamefont {Park}\ \emph {et~al.}(2018)\citenamefont {Park}, \citenamefont {Depeursinge},\ and\ \citenamefont {Popescu}}]{qpireview}%
  \BibitemOpen
  \bibfield  {author} {\bibinfo {author} {\bibfnamefont {Y.}~\bibnamefont {Park}}, \bibinfo {author} {\bibfnamefont {C.}~\bibnamefont {Depeursinge}},\ and\ \bibinfo {author} {\bibfnamefont {G.}~\bibnamefont {Popescu}},\ }\bibfield  {title} {\bibinfo {title} {Quantitative phase imaging in biomedicine},\ }\href@noop {} {\bibfield  {journal} {\bibinfo  {journal} {Nature photonics}\ }\textbf {\bibinfo {volume} {12}},\ \bibinfo {pages} {578} (\bibinfo {year} {2018})}\BibitemShut {NoStop}%
\bibitem [{\citenamefont {Yin}\ \emph {et~al.}(2012)\citenamefont {Yin}, \citenamefont {Kanade},\ and\ \citenamefont {Chen}}]{phasecontrastunderstanding}%
  \BibitemOpen
  \bibfield  {author} {\bibinfo {author} {\bibfnamefont {Z.}~\bibnamefont {Yin}}, \bibinfo {author} {\bibfnamefont {T.}~\bibnamefont {Kanade}},\ and\ \bibinfo {author} {\bibfnamefont {M.}~\bibnamefont {Chen}},\ }\bibfield  {title} {\bibinfo {title} {Understanding the phase contrast optics to restore artifact-free microscopy images for segmentation},\ }\href@noop {} {\bibfield  {journal} {\bibinfo  {journal} {Medical image analysis}\ }\textbf {\bibinfo {volume} {16}},\ \bibinfo {pages} {1047} (\bibinfo {year} {2012})}\BibitemShut {NoStop}%
\bibitem [{\citenamefont {Choi}\ \emph {et~al.}(2007)\citenamefont {Choi}, \citenamefont {Fang-Yen}, \citenamefont {Badizadegan}, \citenamefont {Oh}, \citenamefont {Lue}, \citenamefont {Dasari},\ and\ \citenamefont {Feld}}]{phaseimagingrf1}%
  \BibitemOpen
  \bibfield  {author} {\bibinfo {author} {\bibfnamefont {W.}~\bibnamefont {Choi}}, \bibinfo {author} {\bibfnamefont {C.}~\bibnamefont {Fang-Yen}}, \bibinfo {author} {\bibfnamefont {K.}~\bibnamefont {Badizadegan}}, \bibinfo {author} {\bibfnamefont {S.}~\bibnamefont {Oh}}, \bibinfo {author} {\bibfnamefont {N.}~\bibnamefont {Lue}}, \bibinfo {author} {\bibfnamefont {R.~R.}\ \bibnamefont {Dasari}},\ and\ \bibinfo {author} {\bibfnamefont {M.~S.}\ \bibnamefont {Feld}},\ }\bibfield  {title} {\bibinfo {title} {Tomographic phase microscopy},\ }\href@noop {} {\bibfield  {journal} {\bibinfo  {journal} {Nature methods}\ }\textbf {\bibinfo {volume} {4}},\ \bibinfo {pages} {717} (\bibinfo {year} {2007})}\BibitemShut {NoStop}%
\bibitem [{\citenamefont {Kim}\ \emph {et~al.}(2014)\citenamefont {Kim}, \citenamefont {Zhou}, \citenamefont {Mir}, \citenamefont {Babacan}, \citenamefont {Carney}, \citenamefont {Goddard},\ and\ \citenamefont {Popescu}}]{phaseimagingrf2}%
  \BibitemOpen
  \bibfield  {author} {\bibinfo {author} {\bibfnamefont {T.}~\bibnamefont {Kim}}, \bibinfo {author} {\bibfnamefont {R.}~\bibnamefont {Zhou}}, \bibinfo {author} {\bibfnamefont {M.}~\bibnamefont {Mir}}, \bibinfo {author} {\bibfnamefont {S.~D.}\ \bibnamefont {Babacan}}, \bibinfo {author} {\bibfnamefont {P.~S.}\ \bibnamefont {Carney}}, \bibinfo {author} {\bibfnamefont {L.~L.}\ \bibnamefont {Goddard}},\ and\ \bibinfo {author} {\bibfnamefont {G.}~\bibnamefont {Popescu}},\ }\bibfield  {title} {\bibinfo {title} {White-light diffraction tomography of unlabelled live cells},\ }\href@noop {} {\bibfield  {journal} {\bibinfo  {journal} {Nature Photonics}\ }\textbf {\bibinfo {volume} {8}},\ \bibinfo {pages} {256} (\bibinfo {year} {2014})}\BibitemShut {NoStop}%
\bibitem [{\citenamefont {Cotte}\ \emph {et~al.}(2013)\citenamefont {Cotte}, \citenamefont {Toy}, \citenamefont {Jourdain}, \citenamefont {Pavillon}, \citenamefont {Boss}, \citenamefont {Magistretti}, \citenamefont {Marquet},\ and\ \citenamefont {Depeursinge}}]{phaseimagingrf3}%
  \BibitemOpen
  \bibfield  {author} {\bibinfo {author} {\bibfnamefont {Y.}~\bibnamefont {Cotte}}, \bibinfo {author} {\bibfnamefont {F.}~\bibnamefont {Toy}}, \bibinfo {author} {\bibfnamefont {P.}~\bibnamefont {Jourdain}}, \bibinfo {author} {\bibfnamefont {N.}~\bibnamefont {Pavillon}}, \bibinfo {author} {\bibfnamefont {D.}~\bibnamefont {Boss}}, \bibinfo {author} {\bibfnamefont {P.}~\bibnamefont {Magistretti}}, \bibinfo {author} {\bibfnamefont {P.}~\bibnamefont {Marquet}},\ and\ \bibinfo {author} {\bibfnamefont {C.}~\bibnamefont {Depeursinge}},\ }\bibfield  {title} {\bibinfo {title} {Marker-free phase nanoscopy},\ }\href@noop {} {\bibfield  {journal} {\bibinfo  {journal} {Nature Photonics}\ }\textbf {\bibinfo {volume} {7}},\ \bibinfo {pages} {113} (\bibinfo {year} {2013})}\BibitemShut {NoStop}%
\bibitem [{\citenamefont {Chipman}\ \emph {et~al.}(2018)\citenamefont {Chipman}, \citenamefont {Lam},\ and\ \citenamefont {Young}}]{chipman2018polarized}%
  \BibitemOpen
  \bibfield  {author} {\bibinfo {author} {\bibfnamefont {R.~A.}\ \bibnamefont {Chipman}}, \bibinfo {author} {\bibfnamefont {W.~S.~T.}\ \bibnamefont {Lam}},\ and\ \bibinfo {author} {\bibfnamefont {G.}~\bibnamefont {Young}},\ }\href@noop {} {\emph {\bibinfo {title} {Polarized light and optical systems}}}\ (\bibinfo  {publisher} {CRC press},\ \bibinfo {year} {2018})\BibitemShut {NoStop}%
\bibitem [{\citenamefont {Goldstein}(2017)}]{goldstein2017polarized}%
  \BibitemOpen
  \bibfield  {author} {\bibinfo {author} {\bibfnamefont {D.~H.}\ \bibnamefont {Goldstein}},\ }\href@noop {} {\emph {\bibinfo {title} {Polarized light}}}\ (\bibinfo  {publisher} {CRC press},\ \bibinfo {year} {2017})\BibitemShut {NoStop}%
\bibitem [{\citenamefont {Gil}\ and\ \citenamefont {Ossikovski}(2022)}]{gil2022polarized}%
  \BibitemOpen
  \bibfield  {author} {\bibinfo {author} {\bibfnamefont {J.~J.}\ \bibnamefont {Gil}}\ and\ \bibinfo {author} {\bibfnamefont {R.}~\bibnamefont {Ossikovski}},\ }\href@noop {} {\emph {\bibinfo {title} {Polarized light and the Mueller matrix approach}}}\ (\bibinfo  {publisher} {CRC press},\ \bibinfo {year} {2022})\BibitemShut {NoStop}%
\bibitem [{\citenamefont {Zernike}(1934)}]{zernike1934diffraction}%
  \BibitemOpen
  \bibfield  {author} {\bibinfo {author} {\bibfnamefont {F.}~\bibnamefont {Zernike}},\ }\bibfield  {title} {\bibinfo {title} {Diffraction theory of the knife-edge test and its improved form, the phase-contrast method},\ }\href@noop {} {\bibfield  {journal} {\bibinfo  {journal} {Monthly Notices of the Royal Astronomical Society, Vol. 94, p. 377-384}\ }\textbf {\bibinfo {volume} {94}},\ \bibinfo {pages} {377} (\bibinfo {year} {1934})}\BibitemShut {NoStop}%
\bibitem [{\citenamefont {Zernike}(1942{\natexlab{a}})}]{zernike1942phase}%
  \BibitemOpen
  \bibfield  {author} {\bibinfo {author} {\bibfnamefont {F.}~\bibnamefont {Zernike}},\ }\bibfield  {title} {\bibinfo {title} {Phase contrast, a new method for the microscopic observation of transparent objects},\ }\href@noop {} {\bibfield  {journal} {\bibinfo  {journal} {Physica}\ }\textbf {\bibinfo {volume} {9}},\ \bibinfo {pages} {686} (\bibinfo {year} {1942}{\natexlab{a}})}\BibitemShut {NoStop}%
\bibitem [{\citenamefont {Zernike}(1942{\natexlab{b}})}]{zernike1942phaseII}%
  \BibitemOpen
  \bibfield  {author} {\bibinfo {author} {\bibfnamefont {F.}~\bibnamefont {Zernike}},\ }\bibfield  {title} {\bibinfo {title} {Phase contrast, a new method for the microscopic observation of transparent objects part ii},\ }\href@noop {} {\bibfield  {journal} {\bibinfo  {journal} {Physica}\ }\textbf {\bibinfo {volume} {9}},\ \bibinfo {pages} {974} (\bibinfo {year} {1942}{\natexlab{b}})}\BibitemShut {NoStop}%
\bibitem [{\citenamefont {Burch}\ and\ \citenamefont {Stock}(1942)}]{burch1942phase}%
  \BibitemOpen
  \bibfield  {author} {\bibinfo {author} {\bibfnamefont {C.}~\bibnamefont {Burch}}\ and\ \bibinfo {author} {\bibfnamefont {J.}~\bibnamefont {Stock}},\ }\bibfield  {title} {\bibinfo {title} {Phase-contrast microscopy},\ }\href@noop {} {\bibfield  {journal} {\bibinfo  {journal} {Journal of Scientific Instruments}\ }\textbf {\bibinfo {volume} {19}},\ \bibinfo {pages} {71} (\bibinfo {year} {1942})}\BibitemShut {NoStop}%
\bibitem [{\citenamefont {Martin}(1947)}]{martin1947phase}%
  \BibitemOpen
  \bibfield  {author} {\bibinfo {author} {\bibfnamefont {L.}~\bibnamefont {Martin}},\ }\bibfield  {title} {\bibinfo {title} {Phase-contrast methods in microscopy},\ }\href@noop {} {\bibfield  {journal} {\bibinfo  {journal} {Nature}\ }\textbf {\bibinfo {volume} {159}},\ \bibinfo {pages} {827} (\bibinfo {year} {1947})}\BibitemShut {NoStop}%
\bibitem [{\citenamefont {Nomarski}(1954)}]{nomarski1954dispositif}%
  \BibitemOpen
  \bibfield  {author} {\bibinfo {author} {\bibfnamefont {G.}~\bibnamefont {Nomarski}},\ }\bibfield  {title} {\bibinfo {title} {Dispositif interf{\'e}rentiel {\`a} polarisation pour l’{\'e}tude des objets transparents ou opaques appartenant {\`a} la classe des objets de phase},\ }\href@noop {} {\bibfield  {journal} {\bibinfo  {journal} {Patent FR}\ }\textbf {\bibinfo {volume} {1}},\ \bibinfo {pages} {124} (\bibinfo {year} {1954})}\BibitemShut {NoStop}%
\bibitem [{\citenamefont {Fran{\c{c}}on}(1957)}]{franccon1957polarization}%
  \BibitemOpen
  \bibfield  {author} {\bibinfo {author} {\bibfnamefont {M.}~\bibnamefont {Fran{\c{c}}on}},\ }\bibfield  {title} {\bibinfo {title} {Polarization apparatus for interference microscopy and macroscopy of isotropic transparent objects},\ }\href@noop {} {\bibfield  {journal} {\bibinfo  {journal} {JOSA}\ }\textbf {\bibinfo {volume} {47}},\ \bibinfo {pages} {528} (\bibinfo {year} {1957})}\BibitemShut {NoStop}%
\bibitem [{\citenamefont {Kwon}\ \emph {et~al.}(2020)\citenamefont {Kwon}, \citenamefont {Arbabi}, \citenamefont {Kamali}, \citenamefont {Faraji-Dana},\ and\ \citenamefont {Faraon}}]{kwon2020single}%
  \BibitemOpen
  \bibfield  {author} {\bibinfo {author} {\bibfnamefont {H.}~\bibnamefont {Kwon}}, \bibinfo {author} {\bibfnamefont {E.}~\bibnamefont {Arbabi}}, \bibinfo {author} {\bibfnamefont {S.~M.}\ \bibnamefont {Kamali}}, \bibinfo {author} {\bibfnamefont {M.}~\bibnamefont {Faraji-Dana}},\ and\ \bibinfo {author} {\bibfnamefont {A.}~\bibnamefont {Faraon}},\ }\bibfield  {title} {\bibinfo {title} {Single-shot quantitative phase gradient microscopy using a system of multifunctional metasurfaces},\ }\href@noop {} {\bibfield  {journal} {\bibinfo  {journal} {Nature Photonics}\ }\textbf {\bibinfo {volume} {14}},\ \bibinfo {pages} {109} (\bibinfo {year} {2020})}\BibitemShut {NoStop}%
\bibitem [{\citenamefont {Zhou}\ \emph {et~al.}(2020)\citenamefont {Zhou}, \citenamefont {Zheng}, \citenamefont {Kravchenko},\ and\ \citenamefont {Valentine}}]{zhou2020flat}%
  \BibitemOpen
  \bibfield  {author} {\bibinfo {author} {\bibfnamefont {Y.}~\bibnamefont {Zhou}}, \bibinfo {author} {\bibfnamefont {H.}~\bibnamefont {Zheng}}, \bibinfo {author} {\bibfnamefont {I.~I.}\ \bibnamefont {Kravchenko}},\ and\ \bibinfo {author} {\bibfnamefont {J.}~\bibnamefont {Valentine}},\ }\bibfield  {title} {\bibinfo {title} {Flat optics for image differentiation},\ }\href@noop {} {\bibfield  {journal} {\bibinfo  {journal} {Nature Photonics}\ }\textbf {\bibinfo {volume} {14}},\ \bibinfo {pages} {316} (\bibinfo {year} {2020})}\BibitemShut {NoStop}%
\bibitem [{\citenamefont {Kwon}\ \emph {et~al.}(2018)\citenamefont {Kwon}, \citenamefont {Sounas}, \citenamefont {Cordaro}, \citenamefont {Polman},\ and\ \citenamefont {Al{\`u}}}]{kwon2018nonlocal}%
  \BibitemOpen
  \bibfield  {author} {\bibinfo {author} {\bibfnamefont {H.}~\bibnamefont {Kwon}}, \bibinfo {author} {\bibfnamefont {D.}~\bibnamefont {Sounas}}, \bibinfo {author} {\bibfnamefont {A.}~\bibnamefont {Cordaro}}, \bibinfo {author} {\bibfnamefont {A.}~\bibnamefont {Polman}},\ and\ \bibinfo {author} {\bibfnamefont {A.}~\bibnamefont {Al{\`u}}},\ }\bibfield  {title} {\bibinfo {title} {Nonlocal metasurfaces for optical signal processing},\ }\href@noop {} {\bibfield  {journal} {\bibinfo  {journal} {Physical review letters}\ }\textbf {\bibinfo {volume} {121}},\ \bibinfo {pages} {173004} (\bibinfo {year} {2018})}\BibitemShut {NoStop}%
\bibitem [{\citenamefont {Wang}\ \emph {et~al.}(2022{\natexlab{a}})\citenamefont {Wang}, \citenamefont {He},\ and\ \citenamefont {Luo}}]{wang2022photonic}%
  \BibitemOpen
  \bibfield  {author} {\bibinfo {author} {\bibfnamefont {R.}~\bibnamefont {Wang}}, \bibinfo {author} {\bibfnamefont {S.}~\bibnamefont {He}},\ and\ \bibinfo {author} {\bibfnamefont {H.}~\bibnamefont {Luo}},\ }\bibfield  {title} {\bibinfo {title} {Photonic spin-hall differential microscopy},\ }\href@noop {} {\bibfield  {journal} {\bibinfo  {journal} {Physical Review Applied}\ }\textbf {\bibinfo {volume} {18}},\ \bibinfo {pages} {044016} (\bibinfo {year} {2022}{\natexlab{a}})}\BibitemShut {NoStop}%
\bibitem [{\citenamefont {Shou}\ \emph {et~al.}(2023)\citenamefont {Shou}, \citenamefont {Liu},\ and\ \citenamefont {Luo}}]{shou2023optical}%
  \BibitemOpen
  \bibfield  {author} {\bibinfo {author} {\bibfnamefont {Y.}~\bibnamefont {Shou}}, \bibinfo {author} {\bibfnamefont {J.}~\bibnamefont {Liu}},\ and\ \bibinfo {author} {\bibfnamefont {H.}~\bibnamefont {Luo}},\ }\bibfield  {title} {\bibinfo {title} {When optical microscopy meets all-optical analog computing: A brief review},\ }\href@noop {} {\bibfield  {journal} {\bibinfo  {journal} {Frontiers of Physics}\ }\textbf {\bibinfo {volume} {18}},\ \bibinfo {pages} {42601} (\bibinfo {year} {2023})}\BibitemShut {NoStop}%
\bibitem [{\citenamefont {Zhu}\ \emph {et~al.}(2019)\citenamefont {Zhu}, \citenamefont {Lou}, \citenamefont {Zhou}, \citenamefont {Zhang}, \citenamefont {Huang}, \citenamefont {Li}, \citenamefont {Luo}, \citenamefont {Wen}, \citenamefont {Zhu}, \citenamefont {Gong} \emph {et~al.}}]{zhu2019generalized}%
  \BibitemOpen
  \bibfield  {author} {\bibinfo {author} {\bibfnamefont {T.}~\bibnamefont {Zhu}}, \bibinfo {author} {\bibfnamefont {Y.}~\bibnamefont {Lou}}, \bibinfo {author} {\bibfnamefont {Y.}~\bibnamefont {Zhou}}, \bibinfo {author} {\bibfnamefont {J.}~\bibnamefont {Zhang}}, \bibinfo {author} {\bibfnamefont {J.}~\bibnamefont {Huang}}, \bibinfo {author} {\bibfnamefont {Y.}~\bibnamefont {Li}}, \bibinfo {author} {\bibfnamefont {H.}~\bibnamefont {Luo}}, \bibinfo {author} {\bibfnamefont {S.}~\bibnamefont {Wen}}, \bibinfo {author} {\bibfnamefont {S.}~\bibnamefont {Zhu}}, \bibinfo {author} {\bibfnamefont {Q.}~\bibnamefont {Gong}}, \emph {et~al.},\ }\bibfield  {title} {\bibinfo {title} {Generalized spatial differentiation from the spin hall effect of light and its application in image processing of edge detection},\ }\href@noop {} {\bibfield  {journal} {\bibinfo  {journal} {Physical Review Applied}\ }\textbf {\bibinfo {volume} {11}},\ \bibinfo {pages} {034043} (\bibinfo {year} {2019})}\BibitemShut {NoStop}%
\bibitem [{\citenamefont {He}\ \emph {et~al.}(2020{\natexlab{a}})\citenamefont {He}, \citenamefont {Zhou}, \citenamefont {Chen}, \citenamefont {Shu}, \citenamefont {Luo},\ and\ \citenamefont {Wen}}]{he2020wavelength}%
  \BibitemOpen
  \bibfield  {author} {\bibinfo {author} {\bibfnamefont {S.}~\bibnamefont {He}}, \bibinfo {author} {\bibfnamefont {J.}~\bibnamefont {Zhou}}, \bibinfo {author} {\bibfnamefont {S.}~\bibnamefont {Chen}}, \bibinfo {author} {\bibfnamefont {W.}~\bibnamefont {Shu}}, \bibinfo {author} {\bibfnamefont {H.}~\bibnamefont {Luo}},\ and\ \bibinfo {author} {\bibfnamefont {S.}~\bibnamefont {Wen}},\ }\bibfield  {title} {\bibinfo {title} {Wavelength-independent optical fully differential operation based on the spin--orbit interaction of light},\ }\href@noop {} {\bibfield  {journal} {\bibinfo  {journal} {APL Photonics}\ }\textbf {\bibinfo {volume} {5}} (\bibinfo {year} {2020}{\natexlab{a}})}\BibitemShut {NoStop}%
\bibitem [{\citenamefont {He}\ \emph {et~al.}(2020{\natexlab{b}})\citenamefont {He}, \citenamefont {Zhou}, \citenamefont {Chen}, \citenamefont {Shu}, \citenamefont {Luo},\ and\ \citenamefont {Wen}}]{he2020spatial}%
  \BibitemOpen
  \bibfield  {author} {\bibinfo {author} {\bibfnamefont {S.}~\bibnamefont {He}}, \bibinfo {author} {\bibfnamefont {J.}~\bibnamefont {Zhou}}, \bibinfo {author} {\bibfnamefont {S.}~\bibnamefont {Chen}}, \bibinfo {author} {\bibfnamefont {W.}~\bibnamefont {Shu}}, \bibinfo {author} {\bibfnamefont {H.}~\bibnamefont {Luo}},\ and\ \bibinfo {author} {\bibfnamefont {S.}~\bibnamefont {Wen}},\ }\bibfield  {title} {\bibinfo {title} {Spatial differential operation and edge detection based on the geometric spin hall effect of light},\ }\href@noop {} {\bibfield  {journal} {\bibinfo  {journal} {Optics Letters}\ }\textbf {\bibinfo {volume} {45}},\ \bibinfo {pages} {877} (\bibinfo {year} {2020}{\natexlab{b}})}\BibitemShut {NoStop}%
\bibitem [{\citenamefont {Wang}\ \emph {et~al.}(2022{\natexlab{b}})\citenamefont {Wang}, \citenamefont {He}, \citenamefont {Chen},\ and\ \citenamefont {Luo}}]{wang2022brewster}%
  \BibitemOpen
  \bibfield  {author} {\bibinfo {author} {\bibfnamefont {R.}~\bibnamefont {Wang}}, \bibinfo {author} {\bibfnamefont {S.}~\bibnamefont {He}}, \bibinfo {author} {\bibfnamefont {S.}~\bibnamefont {Chen}},\ and\ \bibinfo {author} {\bibfnamefont {H.}~\bibnamefont {Luo}},\ }\bibfield  {title} {\bibinfo {title} {Brewster differential microscopy},\ }\href@noop {} {\bibfield  {journal} {\bibinfo  {journal} {Applied Physics Letters}\ }\textbf {\bibinfo {volume} {121}} (\bibinfo {year} {2022}{\natexlab{b}})}\BibitemShut {NoStop}%
\bibitem [{\citenamefont {Zhou}\ \emph {et~al.}(2022)\citenamefont {Zhou}, \citenamefont {Wu}, \citenamefont {Zhao}, \citenamefont {Posner}, \citenamefont {Lei}, \citenamefont {Chen}, \citenamefont {Zhang},\ and\ \citenamefont {Liu}}]{zhou2022fourier}%
  \BibitemOpen
  \bibfield  {author} {\bibinfo {author} {\bibfnamefont {J.}~\bibnamefont {Zhou}}, \bibinfo {author} {\bibfnamefont {Q.}~\bibnamefont {Wu}}, \bibinfo {author} {\bibfnamefont {J.}~\bibnamefont {Zhao}}, \bibinfo {author} {\bibfnamefont {C.}~\bibnamefont {Posner}}, \bibinfo {author} {\bibfnamefont {M.}~\bibnamefont {Lei}}, \bibinfo {author} {\bibfnamefont {G.}~\bibnamefont {Chen}}, \bibinfo {author} {\bibfnamefont {J.}~\bibnamefont {Zhang}},\ and\ \bibinfo {author} {\bibfnamefont {Z.}~\bibnamefont {Liu}},\ }\bibfield  {title} {\bibinfo {title} {Fourier optical spin splitting microscopy},\ }\href@noop {} {\bibfield  {journal} {\bibinfo  {journal} {Physical review letters}\ }\textbf {\bibinfo {volume} {129}},\ \bibinfo {pages} {020801} (\bibinfo {year} {2022})}\BibitemShut {NoStop}%
\bibitem [{\citenamefont {Zhou}\ \emph {et~al.}(2021)\citenamefont {Zhou}, \citenamefont {Qian}, \citenamefont {Zhao}, \citenamefont {Tang}, \citenamefont {Wu}, \citenamefont {Lei}, \citenamefont {Luo}, \citenamefont {Wen}, \citenamefont {Chen},\ and\ \citenamefont {Liu}}]{zhou2021two}%
  \BibitemOpen
  \bibfield  {author} {\bibinfo {author} {\bibfnamefont {J.}~\bibnamefont {Zhou}}, \bibinfo {author} {\bibfnamefont {H.}~\bibnamefont {Qian}}, \bibinfo {author} {\bibfnamefont {J.}~\bibnamefont {Zhao}}, \bibinfo {author} {\bibfnamefont {M.}~\bibnamefont {Tang}}, \bibinfo {author} {\bibfnamefont {Q.}~\bibnamefont {Wu}}, \bibinfo {author} {\bibfnamefont {M.}~\bibnamefont {Lei}}, \bibinfo {author} {\bibfnamefont {H.}~\bibnamefont {Luo}}, \bibinfo {author} {\bibfnamefont {S.}~\bibnamefont {Wen}}, \bibinfo {author} {\bibfnamefont {S.}~\bibnamefont {Chen}},\ and\ \bibinfo {author} {\bibfnamefont {Z.}~\bibnamefont {Liu}},\ }\bibfield  {title} {\bibinfo {title} {Two-dimensional optical spatial differentiation and high-contrast imaging},\ }\href@noop {} {\bibfield  {journal} {\bibinfo  {journal} {National science review}\ }\textbf {\bibinfo {volume} {8}},\ \bibinfo {pages} {nwaa176} (\bibinfo {year} {2021})}\BibitemShut {NoStop}%
\bibitem [{\citenamefont {Cordaro}\ \emph {et~al.}(2019)\citenamefont {Cordaro}, \citenamefont {Kwon}, \citenamefont {Sounas}, \citenamefont {Koenderink}, \citenamefont {Al{\`u}},\ and\ \citenamefont {Polman}}]{cordaro2019high}%
  \BibitemOpen
  \bibfield  {author} {\bibinfo {author} {\bibfnamefont {A.}~\bibnamefont {Cordaro}}, \bibinfo {author} {\bibfnamefont {H.}~\bibnamefont {Kwon}}, \bibinfo {author} {\bibfnamefont {D.}~\bibnamefont {Sounas}}, \bibinfo {author} {\bibfnamefont {A.~F.}\ \bibnamefont {Koenderink}}, \bibinfo {author} {\bibfnamefont {A.}~\bibnamefont {Al{\`u}}},\ and\ \bibinfo {author} {\bibfnamefont {A.}~\bibnamefont {Polman}},\ }\bibfield  {title} {\bibinfo {title} {High-index dielectric metasurfaces performing mathematical operations},\ }\href@noop {} {\bibfield  {journal} {\bibinfo  {journal} {Nano letters}\ }\textbf {\bibinfo {volume} {19}},\ \bibinfo {pages} {8418} (\bibinfo {year} {2019})}\BibitemShut {NoStop}%
\bibitem [{\citenamefont {Zhu}\ \emph {et~al.}(2017)\citenamefont {Zhu}, \citenamefont {Zhou}, \citenamefont {Lou}, \citenamefont {Ye}, \citenamefont {Qiu}, \citenamefont {Ruan},\ and\ \citenamefont {Fan}}]{zhu2017plasmonic}%
  \BibitemOpen
  \bibfield  {author} {\bibinfo {author} {\bibfnamefont {T.}~\bibnamefont {Zhu}}, \bibinfo {author} {\bibfnamefont {Y.}~\bibnamefont {Zhou}}, \bibinfo {author} {\bibfnamefont {Y.}~\bibnamefont {Lou}}, \bibinfo {author} {\bibfnamefont {H.}~\bibnamefont {Ye}}, \bibinfo {author} {\bibfnamefont {M.}~\bibnamefont {Qiu}}, \bibinfo {author} {\bibfnamefont {Z.}~\bibnamefont {Ruan}},\ and\ \bibinfo {author} {\bibfnamefont {S.}~\bibnamefont {Fan}},\ }\bibfield  {title} {\bibinfo {title} {Plasmonic computing of spatial differentiation},\ }\href@noop {} {\bibfield  {journal} {\bibinfo  {journal} {Nature communications}\ }\textbf {\bibinfo {volume} {8}},\ \bibinfo {pages} {15391} (\bibinfo {year} {2017})}\BibitemShut {NoStop}%
\bibitem [{\citenamefont {Zhou}\ \emph {et~al.}(2019)\citenamefont {Zhou}, \citenamefont {Qian}, \citenamefont {Chen}, \citenamefont {Zhao}, \citenamefont {Li}, \citenamefont {Wu}, \citenamefont {Luo}, \citenamefont {Wen},\ and\ \citenamefont {Liu}}]{zhou2019optical}%
  \BibitemOpen
  \bibfield  {author} {\bibinfo {author} {\bibfnamefont {J.}~\bibnamefont {Zhou}}, \bibinfo {author} {\bibfnamefont {H.}~\bibnamefont {Qian}}, \bibinfo {author} {\bibfnamefont {C.-F.}\ \bibnamefont {Chen}}, \bibinfo {author} {\bibfnamefont {J.}~\bibnamefont {Zhao}}, \bibinfo {author} {\bibfnamefont {G.}~\bibnamefont {Li}}, \bibinfo {author} {\bibfnamefont {Q.}~\bibnamefont {Wu}}, \bibinfo {author} {\bibfnamefont {H.}~\bibnamefont {Luo}}, \bibinfo {author} {\bibfnamefont {S.}~\bibnamefont {Wen}},\ and\ \bibinfo {author} {\bibfnamefont {Z.}~\bibnamefont {Liu}},\ }\bibfield  {title} {\bibinfo {title} {Optical edge detection based on high-efficiency dielectric metasurface},\ }\href@noop {} {\bibfield  {journal} {\bibinfo  {journal} {Proceedings of the National Academy of Sciences}\ }\textbf {\bibinfo {volume} {116}},\ \bibinfo {pages} {11137} (\bibinfo {year} {2019})}\BibitemShut {NoStop}%
\bibitem [{\citenamefont {Engay}\ \emph {et~al.}(2021)\citenamefont {Engay}, \citenamefont {Huo}, \citenamefont {Malureanu}, \citenamefont {Bunea},\ and\ \citenamefont {Lavrinenko}}]{engay2021polarization}%
  \BibitemOpen
  \bibfield  {author} {\bibinfo {author} {\bibfnamefont {E.}~\bibnamefont {Engay}}, \bibinfo {author} {\bibfnamefont {D.}~\bibnamefont {Huo}}, \bibinfo {author} {\bibfnamefont {R.}~\bibnamefont {Malureanu}}, \bibinfo {author} {\bibfnamefont {A.-I.}\ \bibnamefont {Bunea}},\ and\ \bibinfo {author} {\bibfnamefont {A.}~\bibnamefont {Lavrinenko}},\ }\bibfield  {title} {\bibinfo {title} {Polarization-dependent all-dielectric metasurface for single-shot quantitative phase imaging},\ }\href@noop {} {\bibfield  {journal} {\bibinfo  {journal} {Nano Letters}\ }\textbf {\bibinfo {volume} {21}},\ \bibinfo {pages} {3820} (\bibinfo {year} {2021})}\BibitemShut {NoStop}%
\bibitem [{\citenamefont {Cotrufo}\ \emph {et~al.}(2022)\citenamefont {Cotrufo}, \citenamefont {Arora}, \citenamefont {Singh},\ and\ \citenamefont {Al{\`u}}}]{cotrufo2022dispersion}%
  \BibitemOpen
  \bibfield  {author} {\bibinfo {author} {\bibfnamefont {M.}~\bibnamefont {Cotrufo}}, \bibinfo {author} {\bibfnamefont {A.}~\bibnamefont {Arora}}, \bibinfo {author} {\bibfnamefont {S.}~\bibnamefont {Singh}},\ and\ \bibinfo {author} {\bibfnamefont {A.}~\bibnamefont {Al{\`u}}},\ }\bibfield  {title} {\bibinfo {title} {Dispersion engineered metasurfaces for broadband, high-na, high-efficiency, dual-polarization analog image processing},\ }\href@noop {} {\bibfield  {journal} {\bibinfo  {journal} {arXiv preprint arXiv:2212.03468}\ } (\bibinfo {year} {2022})}\BibitemShut {NoStop}%
\bibitem [{\citenamefont {Silva}\ \emph {et~al.}(2014)\citenamefont {Silva}, \citenamefont {Monticone}, \citenamefont {Castaldi}, \citenamefont {Galdi}, \citenamefont {Al{\`u}},\ and\ \citenamefont {Engheta}}]{silva2014performing}%
  \BibitemOpen
  \bibfield  {author} {\bibinfo {author} {\bibfnamefont {A.}~\bibnamefont {Silva}}, \bibinfo {author} {\bibfnamefont {F.}~\bibnamefont {Monticone}}, \bibinfo {author} {\bibfnamefont {G.}~\bibnamefont {Castaldi}}, \bibinfo {author} {\bibfnamefont {V.}~\bibnamefont {Galdi}}, \bibinfo {author} {\bibfnamefont {A.}~\bibnamefont {Al{\`u}}},\ and\ \bibinfo {author} {\bibfnamefont {N.}~\bibnamefont {Engheta}},\ }\bibfield  {title} {\bibinfo {title} {Performing mathematical operations with metamaterials},\ }\href@noop {} {\bibfield  {journal} {\bibinfo  {journal} {Science}\ }\textbf {\bibinfo {volume} {343}},\ \bibinfo {pages} {160} (\bibinfo {year} {2014})}\BibitemShut {NoStop}%
\bibitem [{\citenamefont {Hwang}\ and\ \citenamefont {Davis}(2016)}]{hwang2016optical}%
  \BibitemOpen
  \bibfield  {author} {\bibinfo {author} {\bibfnamefont {Y.}~\bibnamefont {Hwang}}\ and\ \bibinfo {author} {\bibfnamefont {T.~J.}\ \bibnamefont {Davis}},\ }\bibfield  {title} {\bibinfo {title} {Optical metasurfaces for subwavelength difference operations},\ }\href@noop {} {\bibfield  {journal} {\bibinfo  {journal} {Applied Physics Letters}\ }\textbf {\bibinfo {volume} {109}} (\bibinfo {year} {2016})}\BibitemShut {NoStop}%
\bibitem [{\citenamefont {Davis}\ \emph {et~al.}(2000)\citenamefont {Davis}, \citenamefont {McNamara}, \citenamefont {Cottrell},\ and\ \citenamefont {Campos}}]{davis2000image}%
  \BibitemOpen
  \bibfield  {author} {\bibinfo {author} {\bibfnamefont {J.~A.}\ \bibnamefont {Davis}}, \bibinfo {author} {\bibfnamefont {D.~E.}\ \bibnamefont {McNamara}}, \bibinfo {author} {\bibfnamefont {D.~M.}\ \bibnamefont {Cottrell}},\ and\ \bibinfo {author} {\bibfnamefont {J.}~\bibnamefont {Campos}},\ }\bibfield  {title} {\bibinfo {title} {Image processing with the radial hilbert transform: theory and experiments},\ }\href@noop {} {\bibfield  {journal} {\bibinfo  {journal} {Optics Letters}\ }\textbf {\bibinfo {volume} {25}},\ \bibinfo {pages} {99} (\bibinfo {year} {2000})}\BibitemShut {NoStop}%
\bibitem [{\citenamefont {Davis}\ and\ \citenamefont {Nowak}(2002)}]{davis2002selective}%
  \BibitemOpen
  \bibfield  {author} {\bibinfo {author} {\bibfnamefont {J.~A.}\ \bibnamefont {Davis}}\ and\ \bibinfo {author} {\bibfnamefont {M.~D.}\ \bibnamefont {Nowak}},\ }\bibfield  {title} {\bibinfo {title} {Selective edge enhancement of images with an acousto-optic light modulator},\ }\href@noop {} {\bibfield  {journal} {\bibinfo  {journal} {Applied optics}\ }\textbf {\bibinfo {volume} {41}},\ \bibinfo {pages} {4835} (\bibinfo {year} {2002})}\BibitemShut {NoStop}%
\bibitem [{\citenamefont {Goodman}\ and\ \citenamefont {Sutton}(1996)}]{goodman1996introduction}%
  \BibitemOpen
  \bibfield  {author} {\bibinfo {author} {\bibfnamefont {J.~W.}\ \bibnamefont {Goodman}}\ and\ \bibinfo {author} {\bibfnamefont {P.}~\bibnamefont {Sutton}},\ }\bibfield  {title} {\bibinfo {title} {Introduction to fourier optics},\ }\href@noop {} {\bibfield  {journal} {\bibinfo  {journal} {Quantum and Semiclassical Optics-Journal of the European Optical Society Part B}\ }\textbf {\bibinfo {volume} {8}},\ \bibinfo {pages} {1095} (\bibinfo {year} {1996})}\BibitemShut {NoStop}%
\bibitem [{\citenamefont {Davis}\ \emph {et~al.}(2001)\citenamefont {Davis}, \citenamefont {Smith}, \citenamefont {McNamara}, \citenamefont {Cottrell},\ and\ \citenamefont {Campos}}]{davis2001fractional}%
  \BibitemOpen
  \bibfield  {author} {\bibinfo {author} {\bibfnamefont {J.~A.}\ \bibnamefont {Davis}}, \bibinfo {author} {\bibfnamefont {D.~A.}\ \bibnamefont {Smith}}, \bibinfo {author} {\bibfnamefont {D.~E.}\ \bibnamefont {McNamara}}, \bibinfo {author} {\bibfnamefont {D.~M.}\ \bibnamefont {Cottrell}},\ and\ \bibinfo {author} {\bibfnamefont {J.}~\bibnamefont {Campos}},\ }\bibfield  {title} {\bibinfo {title} {Fractional derivatives—analysis and experimental implementation},\ }\href@noop {} {\bibfield  {journal} {\bibinfo  {journal} {Applied Optics}\ }\textbf {\bibinfo {volume} {40}},\ \bibinfo {pages} {5943} (\bibinfo {year} {2001})}\BibitemShut {NoStop}%
\bibitem [{\citenamefont {Gupta}\ \emph {et~al.}(2015)\citenamefont {Gupta}, \citenamefont {Ghosh},\ and\ \citenamefont {Banerjee}}]{gupta2015wave}%
  \BibitemOpen
  \bibfield  {author} {\bibinfo {author} {\bibfnamefont {S.~D.}\ \bibnamefont {Gupta}}, \bibinfo {author} {\bibfnamefont {N.}~\bibnamefont {Ghosh}},\ and\ \bibinfo {author} {\bibfnamefont {A.}~\bibnamefont {Banerjee}},\ }\href@noop {} {\emph {\bibinfo {title} {Wave optics: Basic concepts and contemporary trends}}}\ (\bibinfo  {publisher} {CRC Press},\ \bibinfo {year} {2015})\BibitemShut {NoStop}%
\bibitem [{\citenamefont {Modak}\ \emph {et~al.}(2021)\citenamefont {Modak}, \citenamefont {Athira}, \citenamefont {Singh},\ and\ \citenamefont {Ghosh}}]{modak2021generalized}%
  \BibitemOpen
  \bibfield  {author} {\bibinfo {author} {\bibfnamefont {N.}~\bibnamefont {Modak}}, \bibinfo {author} {\bibfnamefont {B.}~\bibnamefont {Athira}}, \bibinfo {author} {\bibfnamefont {A.~K.}\ \bibnamefont {Singh}},\ and\ \bibinfo {author} {\bibfnamefont {N.}~\bibnamefont {Ghosh}},\ }\bibfield  {title} {\bibinfo {title} {Generalized framework of weak-value amplification in path interference of polarized light for the enhancement of all possible polarization anisotropy effects},\ }\href@noop {} {\bibfield  {journal} {\bibinfo  {journal} {Physical Review A}\ }\textbf {\bibinfo {volume} {103}},\ \bibinfo {pages} {053518} (\bibinfo {year} {2021})}\BibitemShut {NoStop}%
\bibitem [{\citenamefont {Pal}\ \emph {et~al.}(2016)\citenamefont {Pal}, \citenamefont {Banerjee}, \citenamefont {Chandel}, \citenamefont {Bag}, \citenamefont {Majumder},\ and\ \citenamefont {Ghosh}}]{pal2016tunable}%
  \BibitemOpen
  \bibfield  {author} {\bibinfo {author} {\bibfnamefont {M.}~\bibnamefont {Pal}}, \bibinfo {author} {\bibfnamefont {C.}~\bibnamefont {Banerjee}}, \bibinfo {author} {\bibfnamefont {S.}~\bibnamefont {Chandel}}, \bibinfo {author} {\bibfnamefont {A.}~\bibnamefont {Bag}}, \bibinfo {author} {\bibfnamefont {S.~K.}\ \bibnamefont {Majumder}},\ and\ \bibinfo {author} {\bibfnamefont {N.}~\bibnamefont {Ghosh}},\ }\bibfield  {title} {\bibinfo {title} {Tunable spin dependent beam shift by simultaneously tailoring geometric and dynamical phases of light in inhomogeneous anisotropic medium},\ }\href@noop {} {\bibfield  {journal} {\bibinfo  {journal} {Scientific reports}\ }\textbf {\bibinfo {volume} {6}},\ \bibinfo {pages} {39582} (\bibinfo {year} {2016})}\BibitemShut {NoStop}%
\bibitem [{\citenamefont {Efron}(1994)}]{efron1994spatial}%
  \BibitemOpen
  \bibfield  {author} {\bibinfo {author} {\bibfnamefont {U.}~\bibnamefont {Efron}},\ }\href@noop {} {\emph {\bibinfo {title} {Spatial light modulator technology: materials, devices, and applications}}},\ Vol.~\bibinfo {volume} {47}\ (\bibinfo  {publisher} {CRC press},\ \bibinfo {year} {1994})\BibitemShut {NoStop}%
\end{thebibliography}%
\end{document}


\maketitle
\section{Theoretical framework of the Signum phase mask differential microscopy}
The proposed spatial differentiation scheme is achieved by placing a Signum phase mask [41] in the Fourier plane of a conventional $4-f$ imaging setup [41]. The Signum phase mask modulates the spatial frequencies ($k_x, k_y$) of the input electric field $E_{in}(k_x,k_y)$, and the the modulated electric field has the form
\begin{equation}
\begin{split}
    \vec{E}(k_x, k_y) = \Vec{E_{in}}(k_x, k_y) \times Sgn(k_x,k_y)
\end{split}
\label{eq1.1}
\end{equation}
The resultant output electric field can be calculated by taking the Fourier transformation of the field at the Fourier plane. Using the Convolution, the expression is found to be
\begin{equation}
\begin{split}
    E_{out}(x',y') \propto &\int \frac{dE_{in}(x,y)}{dy} \ln |y'-y|\,dy
\end{split}
\label{eq1.2}
\end{equation}
In our study, we have considered a $1D$ Signum function as a phase mask ($f(k_y) = \{1: k_y >0, -1: k_y$$<0, 0: k_y=0\}$), which leads to a $1D$ spatial differentiator along the direction of the Signum function (in the $y$ direction). A more complicated Signum phase mask can be judiciously chosen to achieve $2D$ isotropic spatial differentiation. However, as our aim is to demonstrate a simple multi-functional differential microscopy, the phase mask is decided accordingly. From the obtained expression, it is evident that the output intensity is directly proportional to the derivative of the input electric field, i.e., $E_{out}\propto\frac{dE_{in}}{dy}$, which is the signature of typical differential microscopy [29]. Now if the incident electric field $\vec{E}_{in}$ or the input object contains finite jumps or discontinuities (at $y=a_1,a_2,a_3,\ldots,a_n$) in terms of either amplitude, phase, or polarization, then the output electric field obtained in Eq.\eqref{eq1.2} takes the following form. 
\begin{equation}
\begin{split}
    &E_{out}(x',y')\\ &\propto \int {(\delta(a_1) + \delta(a_2) +\delta(a_3)+.....+\delta (a_n)) \ln |y'-y|}\,dy \\
    &\propto \ln |y'-a_1| + \ln |y'-a_2| + \ln |y'-a_3| +......+ \ln |y'-a_n|
\end{split}
\label{eq1.3}
\end{equation}
From the above expression, it is evident that there is an intensity localization or enhancement at the point of field discontinuities, which leads to the formation of differential-contrast images for arbitrary phase, polarization, and amplitude objects. The existing differential microscopy systems have primarily been utilized for imaging phase objects. However, the polarization properties of an object have not been considered as an intrinsic contrast mechanism in the domain of differential microscopy, despite the fact that polarization imaging has been an important tool in label-free microscopy with potential applications in various disciplines. The reason behind this is most of the differential microscopy methods utilize the polarization degree of freedom to obtain the spatial differentiation, hence making it impossible to probe the polarization properties of the imaging specimen in the same experimental embodiment. Our proposed spatial differentiator possesses no dependence on polarization, which enables us to consider the vectorial nature of the input electric field. As a result, alongside the conventional differential system, quantitative polarization-resolved and polarization gradient imaging are established in a single experimental embodiment. To understand the polarization gradient imaging, let us consider a generalized electric field, which has the form
\begin{equation}
\begin{split}
    \vec{E}_{obj}(x,y) = \begin{pmatrix} E_1(x,y)\\E_2(x,y) \end{pmatrix} e^{i\frac{2\pi}{\lambda} (n^{r}(x,y)+in^{i}(x,y))z}\\
    = E_0(x,y) e^{i\delta(x,y)}\begin{pmatrix} \hat{E_1}(x,y) e^{i\delta_x(x,y)}\\ \hat{E_2}(x,y) e^{i\delta_y(x,y)} \end{pmatrix}
\end{split}
\label{eq1.4}
\end{equation}
The above expressions describe a vector electric field propagating in the ${z}$ direction after experiencing an arbitrary object. While $E_0(x,y)$, and $e^{i\delta_0(x,y)}$ describe the spatial variations of the absolute amplitude and phase, respectively, the polarization properties are encoded in the Jones vector [43]. As we have discussed above, the conventional differential imaging systems and other methods probe only the spatial gradient of the $E_0(x,y)$, and $e^{i\delta_0(x,y)}$, yet no information regarding the transformation of the vectorial nature and its gradient $\begin{pmatrix}\hat{E_1}(x,y) e^{i\delta_x(x,y)}\\\hat{E_2}(x,y) e^{i\delta_y(x,y)} \end{pmatrix}$ are investigated. It is well known that the output intensity in a differential imaging system is directly proportional to the gradient of amplitude or phase $I_{out} \propto dE_0(x,y),d\delta_0(x,y)$, which leads to the formation of amplitude or phase contrast images. In our differential imaging scheme, as we have a $1D$ spatial differentiator, the output resultant intensity pattern has the form $I_{out} \propto \frac{dE_0(x,y)}{dy}, \frac{d\delta_0(x,y)}{dy}$. Alongside this, the output intensity also probes the spatial gradient of the polarization properties throughout a specimen, which requires detailed discussion as follows.
\par
The polarization properties of an object are primarily probed and quantified through various polarization anisotropy effects such as linear/circular diattenuation, retardance, and depolarization [12, 43]. The spatial map of these anisotropic parameters throughout a specimen reveals its essential structural and morphological information and has been widely adopted in biological and clinical research. We define the generalized polarization anisotropy parameter $\alpha$ in terms of the difference in refractive index (RI), $n$ between orthogonal linear or circular polarizations as, $2\alpha=\frac{2\pi}{\lambda} (n_{L\slash C}^{r\slash i}-n_{L'\slash C'}^{r\slash i})t$ [44]. Here, $t$ is the path length, and the superscripts ($r\slash i$) correspond to real/imaginary parts of RI, and the subscripts ($L\slash C$) and ($L'\slash C'$ ) represent linear/circular polarizations and their orthogonal states, respectively. The real/ imaginary parts of RI generate amplitude/ phase anisotropies and are associated with retardance/ diattenuation effects, respectively. The objective of our spatial differential scheme is to quantify the spatial gradient of these polarization anisotropic parameters, and consequently produce contrast images for pure polarization objects. To demonstrate this, we have considered the circular and linear retardance effects, which are typically exhibited by natural objects with biological origin. The polarization transformation matrix describing the circular retardance ($\alpha=\frac{\pi}{\lambda} (n_{C}^{i}-n_{C'}^{i})t$) effect, can be described by the Jones vector $J_r=\begin{pmatrix}
    \cos{\alpha} & \sin{\alpha}\\
    -\sin{\alpha} & \cos{\alpha}
\end{pmatrix}$.
The incoming electric field modified by the media has the form $\vec{E}_{out}=J_r\times \vec{E}_{in}$, and the consequent change or the gradient in the polarization properties can be formulated as 
\begin{equation}
\begin{pmatrix}
    E_x^{'}\\
    E_y^{'}
\end{pmatrix}
    =\begin{pmatrix}
    \cos{\alpha} & \sin{\alpha}\\
    -\sin{\alpha} & \cos{\alpha}
    \end{pmatrix}
    \times \begin{pmatrix}
        E_x\\
        E_y
    \end{pmatrix},
    \Delta \vec{E}= \begin{pmatrix}
        E_x (\cos{\alpha}-1)+E_y\sin{\alpha}\\
        -E_x \sin{\alpha}+E_y(\cos{\alpha}-1)
    \end{pmatrix}
    \label{eq1.5}
\end{equation}
We have already discussed that the output electric field explicitly depends on the electric field gradient $E_{out}\propto\Delta E$, which is observed in the measured intensity signal. In this scenario (Eq. \eqref{eq1.5}), we need to record the polarization Stokes vector [43] corresponding to the output electric field $\vec{E}_{out}$ to probe and quantify the gradient of the anisotropic parameter (circular retardance, i.e., optical rotation in a linearly polarized light). The Stokes vector for an input $\Vec{x}$ polarized electric field can be obtained as
\begin{equation}
    \begin{pmatrix}
          S_0\\
          S_1\\
          S_2\\
         S_3\\
    \end{pmatrix}
    = \begin{pmatrix}
          <E_xE_x^*>+<E_yE_y^*>\\
          <E_xE_x^*>-<E_yE_y^*>\\
          <E_xE_y^*>+<E_yE_x^*>\\
         <E_xE_y^*>-<E_yE_x^*>\\
    \end{pmatrix}
    = \begin{pmatrix}
        2(1-\cos{\alpha})\\
        2 \cos{\alpha} (\cos{\alpha}-1)\\
        2 \sin{\alpha} (\cos{\alpha}-1)\\
        0        
    \end{pmatrix}
\end{equation}\\
The corresponding results are given in the main text (Fig. 3b), where intensity localization around the edge of a circular object is observed in the Stokes vector elements. The spatial gradient of the rotation parameter can be quantified by measuring the rotation of the output electric field $(\alpha)=1/2\times \arctan(S2/S1)$ [43]. Similar formalism can also be incorporated for arbitrary polarization effects, e.g., linear retardance. Similar to Eq. \eqref{eq1.5}, here, the output Stokes parameters for an incoming $+45\deg$ linearly polarized can be expressed as
\begin{equation}
 \begin{pmatrix}
    E_x^{'}\\
    E_y^{'}
\end{pmatrix}
    =\begin{pmatrix}
    e^{\alpha/2} & 0\\
    0 & e^{-\alpha/2}
    \end{pmatrix}
    \times \begin{pmatrix}
        E_x\\
        E_y
    \end{pmatrix}, 
    \Delta E= \begin{pmatrix}
        E_x (e^{\alpha/2}-1)\\
        E_y (e^{-\alpha/2}-1)
    \end{pmatrix}
\end{equation}
\begin{equation}
    \begin{pmatrix}
          S_0\\
          S_1\\
          S_2\\
         S_3\\
    \end{pmatrix}
    = \begin{pmatrix}
        (1-\cos{\alpha/2})\\
         0\\
         \cos{\alpha/2}(1-\cos{\alpha/2})\\
         \sin{\alpha/2}(1-\cos{\alpha/2})  \end{pmatrix}   
\end{equation}
The spatial gradient in the linear retardance parameter can be quantified by measuring the ellipticity of the output electric field $(\alpha)=1/2\times \arctan(S3/S0)$. Although recording the Stokes parameter for an appropriate input electric field can probe the polarization gradient of a particular polarization effect, it is not adequate where multiple polarization effects occur simultaneously. In such cases, one needs to record the full $4\times4$ Mueller matrix, and through judicious Muller matrix analysis such as polar and differential decomposition, complete information on the exhibited polarization effects can be extracted.
\section{Differential image of pure polarization and amplitude objects}
According to the proposed differential scheme, intensity localization occurs at the discontinuity points in the input electric field. The formation of such differential contrast images for pure phase objects is explicitly demonstrated in the main text. Besides phase objects, the differential image of pure amplitude and polarization objects can be produced in our experimental setup owing to its functionality for general vector electric fields. A circular polarization object exhibiting linear retardance is prepared and placed in an isotropic background. As expected, this polarization gradient leads to the localized intensity at the edge of the circle, which is observed in the Stokes vector of the output image. The spatial polarization gradient is also quantified using standard polarization algebra, as discussed in the previous section.

\begin{figure}[H]
      \centering
      \includegraphics[scale=0.6]{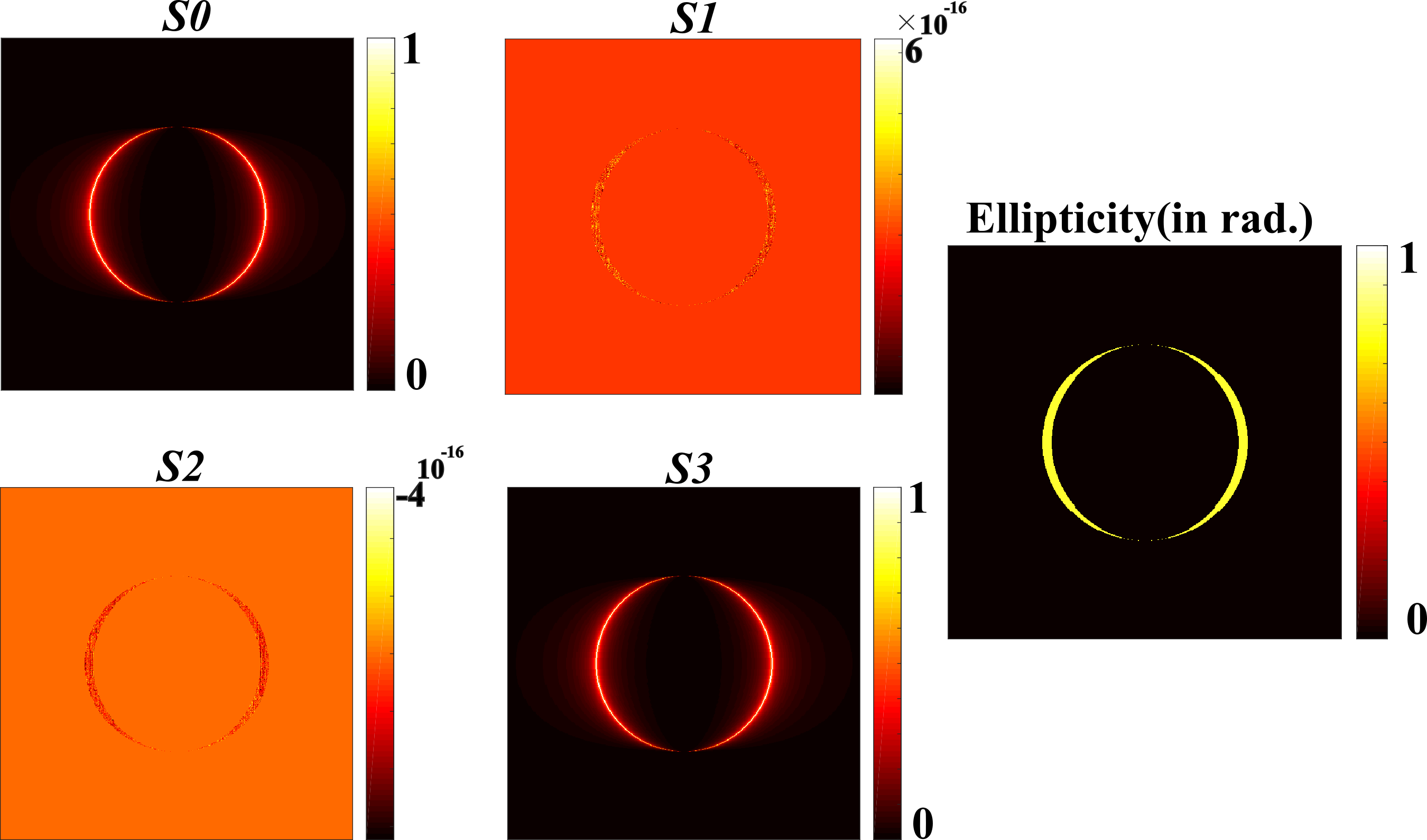}
      \caption{Quantitative polarization gradient imaging. The differential image of a circle-shaped pure polarization object is presented, where the polarization inside a circle acquired slight ellipticity (due to the linear retardance effect) with respect to the background's polarization. The discontinuity in the polarization leads to an intensity localization around the circumference of the circle (Stokes parameter I). The corresponding Stokes parameters and the calculated polarization change in the 4th Stokes parameter $S3$ demonstrate the quantitative differential polarization anisotropy imaging.}
      \label{figS1}
\end{figure}
\par
\begin{figure}[H]
    \centering
    \includegraphics[width=0.8\textwidth]{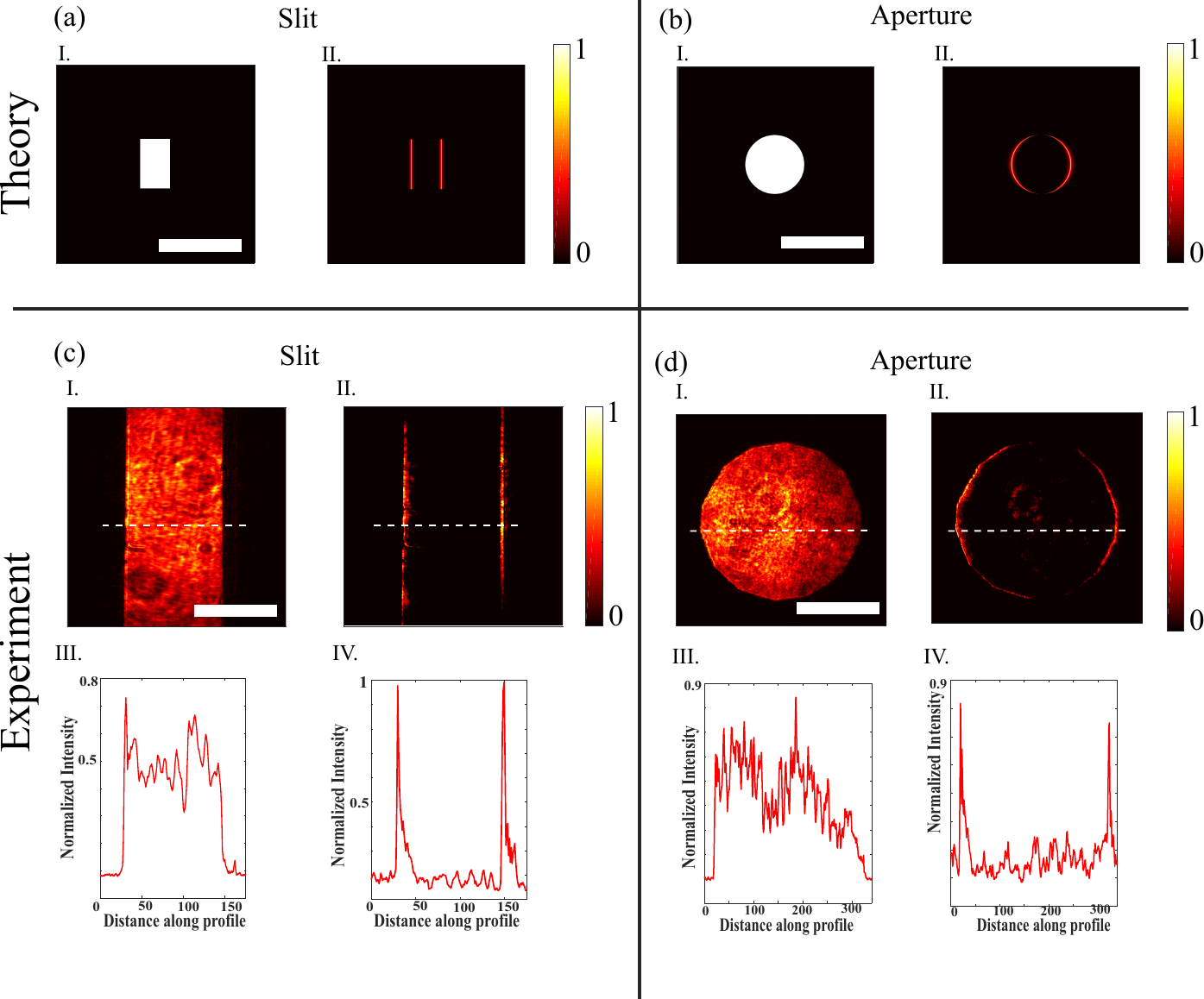}
     \caption{Theoretical ((a), (b)) and experimental ((c), (d)) demonstration of Signum phase mask differential (SPMD) imaging of amplitude objects. A regular rectangular slit ((a), (c)) and a circular aperture ((b), (d)) are considered as imaging targets. (a) Theoretically obtained bright-field (I.) and SPMD (II.) image of a rectangular slit. (b) Theoretically obtained bright-field (I.) and SPMD (II.) image of a circular aperture. (c) Experimentally obtained bright-field (I.) and SPMD (II.) image of a rectangular slit. (d) Experimentally obtained bright-field (I.) and SPMD (II.) image of a circular aperture. The intensity profiles along the white dashed lines of experimentally obtained bright-field (III.) and SPMD (IV.) images are presented.}
     \label{figs2}
\end{figure}

Besides the phase objects (as demonstrated in the main text), we also image regular amplitude objects in the demonstrated SPMD imaging method. A vertical rectangular slit and a circular aperture are used as imaging targets (see Figure S2 \ref{figs2}). As described in Eq. (S.3) and Eq. (S.4) of the above section, the differential images exhibit intensity localization at the edges of the objects, confirming the proposed full-field spatial differentiator.
\newpage
\section{Quantitative phase gradient imaging using the SPMD scheme}
As discussed in previous sections, the proposed differential scheme senses the local derivative of the input electric field, and the intensity of the output image is directly proportional to the spatial gradient of the vectorial transmittance function of the input object. Although the obtained expressions for the resultant intensity (Eq. S.3) imply a blowup of intensity values, a proportionality relation between the electric field gradient and the consequent localized intensity value can be established. This can be understood by considering the non-ideal behaviour of the step function in a real experimental scenario and its corresponding derivative (non-idealized delta function), which yields a finite value of the localized intensity at the output image plane. Hence, a proper calibration technique can be utilized prior to extracting the quantitative phase gradient information of a specimen, which is demonstrated next. In this way, the proposed spatial differentiator can be seen as a semi-quantitative phase gradient imaging.
\par
A circular phase object is considered, where the phase difference between the object and background is increased gradually, and its consequent effect in the localized intensity at the edge of the circle is studied. As expected, the localized intensity at the circumference shows an increment with the spatial phase gradient (Fig. S3 (b)). A monotonic increase in the intensity value is observed for the phase difference from $0$ to $\pi$, and the exact path is traversed back when the phase difference is varied from $\pi$ to $2\pi$ (main text). Hence, a direct relation between the phase gradient and the magnitude of the localized intensity at that particular point can be derived. Here, the obtained curve (Fig. S3 (a)) is fitted with a linear function, which provides the direct relation between the intensity magnitude and the local phase gradient. However, arbitrary polynomial functions can also be used according to the requirement. Experimentally, the system is calibrated by preparing a phase object using the SLM, and the grey level of the phase object is constantly increased. As discussed in the method section of the main text, the phase, or the phase gradient (here), gets increased with the grey levels and exhibits a linear behaviour for a particular range. Hence the magnitude of the localized intensity follows similar behaviour with the gradual increment of the grey levels (Fig. S3 (c)). By fitting the obtained curve with a linear function, a one-to-one relation between the phase difference and the localized intensity value is derived (Fig. S3(c)), which is then used to extract the phase gradient information of unknown samples.
\begin{figure}[H]
    \centering
    \includegraphics[width=\textwidth]{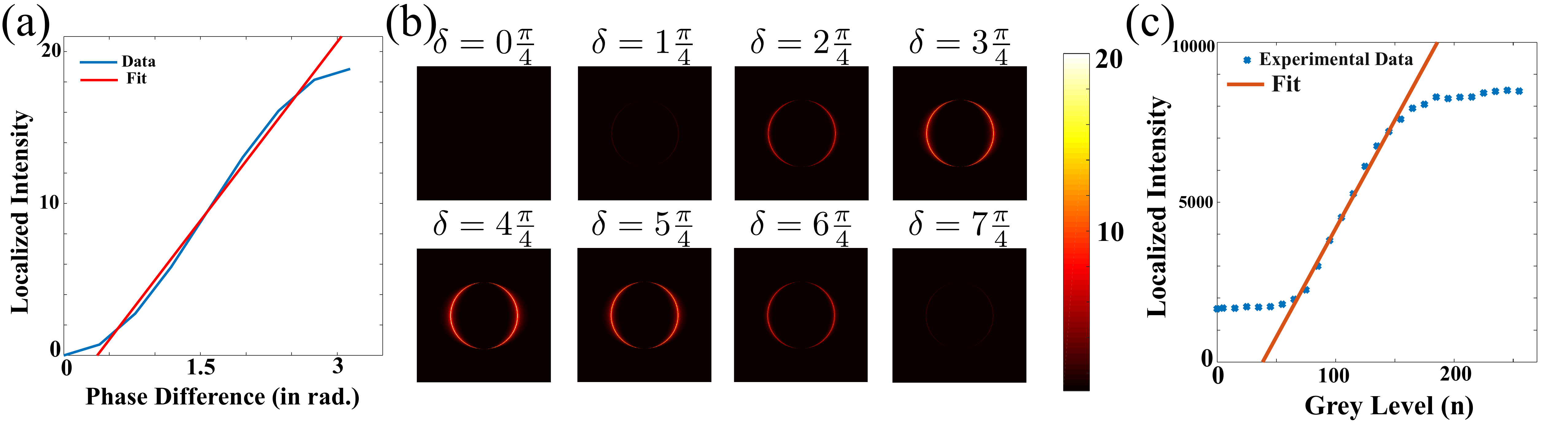}
     \caption{Theoretical variation of intensity localization in SPMD images of a circular phase object with changing phase difference $\delta$ (in the units of $\frac{\pi}{4}$ rad). The colour bar indicates the value of localized intensity (arbitrary unit) at the edge of the circles.}
     \label{figs3}
\end{figure}
\section{Polarization resolved differential microscopy}
The proposed spatial differentiator does not require polarized light to obtain the spatial differentiation, which enables the utilization of polarization degree of freedom as an additional contrast mechanism in the realm of differential microscopy. As a result, we have demonstrated two new kinds of microscopy techniques named polarization gradient imaging, and polarization-resolved differential microscopy. While the detailed theoretical modeling and results for the polarization gradient imaging are discussed in Sec. S.1, some additional results further illustrating the advantages achieved through the merging between polarization and differential imaging are presented below. In the main text, it is already shown that the polarization-resolved measurements can further enhance the contrast of differential images of phase-polarization objects by significantly reducing the unwanted background signal. Thus polarization-resolved differential microscopy can also be advantageous for imaging natural objects possessing intrinsic polarization anisotropic effects. Cross-polarized differential images of potato starch granules (Fig. S4 (a)), and radial liquid crystal droplets (Fig. S4(b)) are recorded, which reveal the internal structural configuration or arrangement of the anisotropy axis. We then go ahead to record the differential images and probe the polarization properties of a phase-polarization object prepared by the SLM. An object with the shape of a cartoon iceman is used as an imaging target, which, besides being a phase object, exhibits multiple polarization effects. Owing to the properties of the differential imaging system, the intensity gets localized at the edge of the object, revealing its structure (Fig. S4(c)). More importantly, the complete polarization properties of the object are measured via recording the $4\times4$ Mueller matrix [43]. The various polarization properties, such as linear/circular diattenuation/retardance, are encoded in the different characteristic Mueller matrix elements. We also note that the spatial variation of the polarization anisotropy parameters can be quantified by utilizing the Mueller matrix decomposition techniques. This way, the proposed SPMD imaging scheme can produce simultaneous quantitative polarization and differential phase images of an arbitrary natural object.
\begin{figure}[H]
    \centering
    \includegraphics[width=0.8\textwidth]{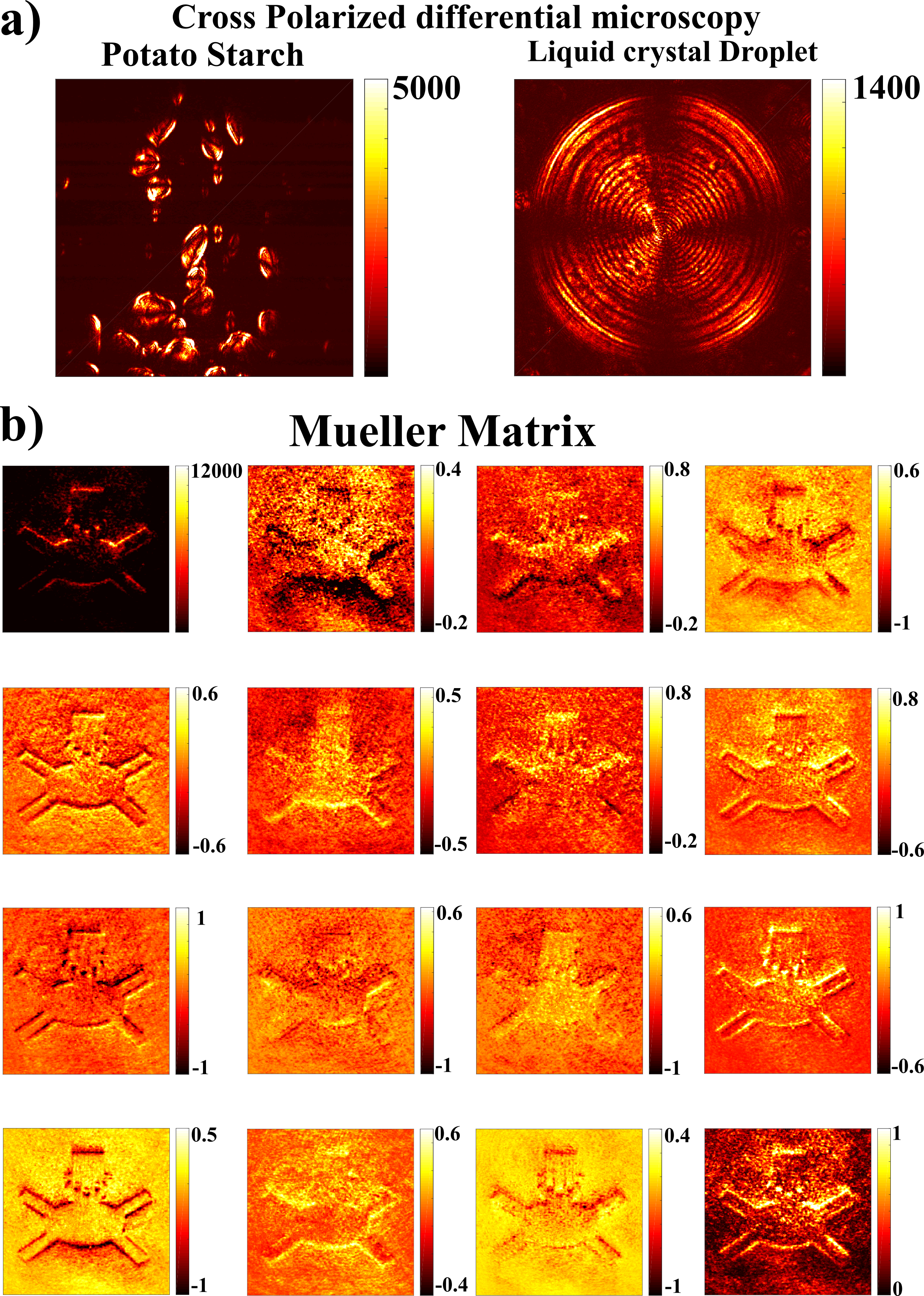}
     \caption{Simultaneous phase and polarization differential imaging. (a) Cross-polarized differential images of potato starch granules and a radial liquid crystal droplet are shown. The high contrast polarization resolved image with lobes-like intensity patterns reveals the internal morphology of the starch granules and the arrangement of the liquid crystals. (b) The $4\times 4$  differential polarization Mueller matrix of a phase-polarization object (Snowman cartoon) is presented, which, alongside producing the differential images, probes and quantifies the complete polarization properties of the object. The non-zero magnitude of the different elements of the Mueller matrix describes the presence of multiple polarization effects, and through appropriate decomposition methods, spatial variation of the polarization anisotropy parameters can be derived.}
     \label{figs4}
\end{figure}

\section{Broadband SPMD imaging}
Another advantage of the proposed SPMD imaging technique is its independence on the wavelength of the incident illumination. As the obtained spatial differentiator is intrinsically independent of the illumination wavelength, the demonstrated differential microscopy can be utilized at arbitrary wavelengths and even for broad-band wavelength ranges to produce phase and polarization contrast images. In this regard, it is worth mentioning that the meta-surface systems used for spatial differentiation usually come with the trade-off of working wavelength regions, which hinders its incorporation in achromatic microscopy. However, our proposed SPMD scheme can be easily integrated with the conventional existing bright-field microscope to produce contrast images of natural phase/polarization objects at arbitrary wavelengths. To demonstrate the broad-band ability of our imaging system, differential images of phase objects are captured using a laser of wavelength $473nm$. The produced high-contrast images of the phase objects provide clear evidence of the working capability of our imaging system at an arbitrary wavelength (see Fig. \ref{figs3}).

\begin{figure}[H]
    \centering
    \includegraphics[width=0.5\textwidth]{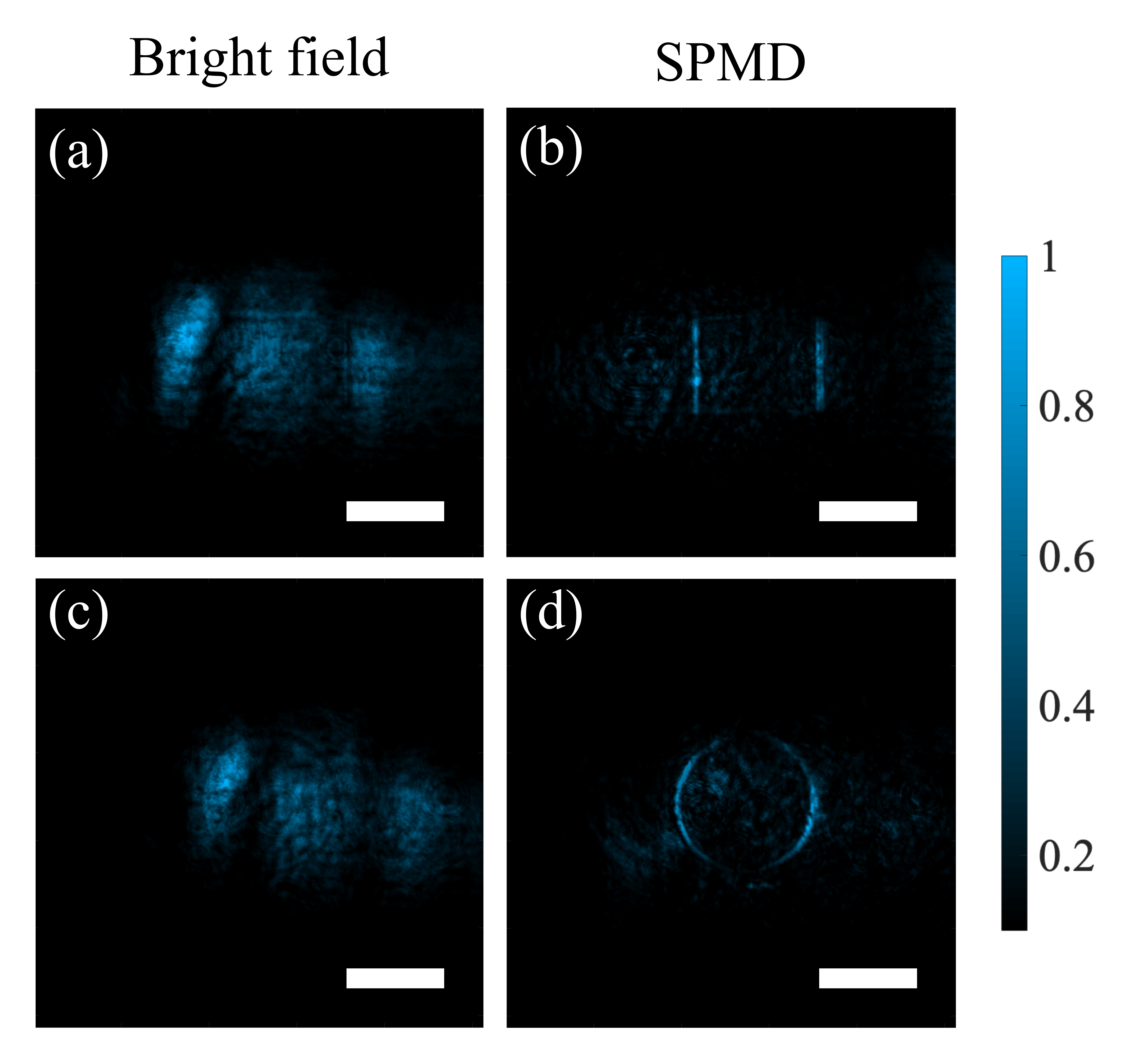}
     \caption{SPMD imaging using $473nm$ diode laser. The bright field ((a), (c)) and differential ((b), (d)) images are obtained for rectangular ((a), (b)), and circular ((c), (d)) phase objects produced by the SLM. The diode laser has some spatially structured emission, as evident in the bright field images. However, the differential images produce high-contrast images. White scale bars indicate a length of $2400\mu m$. The colour bar indicates normalized intensity.}
     \label{figs3}
\end{figure}
\bibliographystyle{ieeetr} 
\bibliography{apssamp}